\documentclass[12pt]{article}
\usepackage[margin=1in]{geometry}
\usepackage{lscape,xcolor}
\usepackage{graphicx}
\usepackage{subfig}
\usepackage{amsfonts}
\usepackage{amsthm}
\usepackage{amsmath}
\usepackage{amssymb}
\usepackage{dsfont}
\usepackage{enumerate}

\usepackage{soul}

\usepackage{natbib}
\usepackage{multirow}
\usepackage{booktabs}
\usepackage{color}
\usepackage{float}
\usepackage{algorithm}
\usepackage{ragged2e}
\usepackage{setspace}
\usepackage[]{algpseudocode}
\usepackage[toc,page]{appendix}

\makeatletter
\renewcommand*\env@matrix[1][*\c@MaxMatrixCols c]{%
  \hskip -\arraycolsep
  \let\@ifnextchar\new@ifnextchar
  \array{#1}}
  \makeatother

\newcommand{\E}{\mathbb{E}}
\newcommand{\Var}{\text{Var}}
\newcommand{\Cov}{\text{Cov}}

\newcommand{\btheta}{\boldsymbol{\theta}}

\newcommand{\D}{\mathbf{D}}

\newcommand{\U}{\mathbf{U}}

\newcommand{\Y}{\mathbf{Y}}
\newcommand{\Z}{\mathbf{Z}}

\newcommand{\mbG}{\mathbf{G}}

\newtheorem*{defn*}{Definition}

\newtheorem{assumption}{Assumption}

\newtheoremstyle{iremark}
  {\topsep}   % ABOVESPACE
  {\topsep}   % BELOWSPACE
  {\upshape}  % BODYFONT
  {0pt}       % INDENT (empty value is the same as 0pt)
  {\itshape}  % HEADFONT
  {.}         % HEADPUNCT
  {5pt plus 1pt minus 1pt} % HEADSPACE
  {\thmname{#1}\thmnumber{ \itshape#2}\thmnote{ (#3)}} % CUSTOM-HEAD-SPEC

\theoremstyle{iremark}

\definecolor{candypink}{rgb}{0.89, 0.44, 0.48}
\definecolor{hgreen}{rgb}{0.21, 0.37, 0.23}

\usepackage{hyperref}
\hypersetup{
    colorlinks = true,
    linkbordercolor = white,
    citecolor = blue,
}

\usepackage{natbib}
\begin{document}

\title{Design of egocentric network-based studies to estimate causal effects under interference}

\author{Junhan Fang\footnote{Center for Methods in Implementation and Prevention Science, Yale School of Public Health, Yale University, New Haven, CT}, Donna Spiegelman\footnotemark[1] \footnote{\baselineskip=10pt Department of Biostatistics, Yale School of Public Health, Yale University, New Haven, CT}, Ashley Buchanan\footnote{ Department of Pharmacy Practice, College of Pharmacy, University of Rhode Island, Kingston, RI}, Laura Forastiere\footnotemark[1] \footnotemark[2] 
}

\pagenumbering{gobble}

\maketitle

%\volume{64}
%\pubyear{2008}
%\artmonth{December}

%  The \doi command is where the DOI for your paper would be placed should it
%  be published.  Again, if you make one up and stick it here, it means 
%  nothing!

%\doi{10.1111/j.1541-0420.2005.00454.x}

%  This label and the label ``lastpage'' are used by the \pagerange
%  command above to give the page range for the article.  You may have 
%  to process the document twice to get this to match up with what you 
%  expect.  When using the referee option, this will not count the pages
%  with tables and figures.  

\label{firstpage}

%  put the summary for your paper here

\begin{abstract}
Many public health interventions are conducted in settings where individuals are connected to one another and the intervention assigned to randomly selected individuals may spill over to other individuals they are connected to. In these spillover settings, the effects of such interventions can be quantified in several ways.  The average individual effect measures the intervention effect among those directly treated, while the spillover effect measures the effect among those connected to those directly treated. In addition, the overall effect measures the average intervention effect across the study population, over those directly treated along with those to whom the intervention spills over but who are not directly treated. Here, we develop methods for study design with the aim of estimating individual, spillover, and overall effects. 
In particular, we consider an egocentric network-based randomized design in which a set of index participants is recruited from the population and randomly assigned to treatment,
while data are also collected from their untreated network members. We use the potential
outcomes framework to define two clustered regression modeling approaches and clarify the underlying assumptions required to identify and estimate causal effects. We then develop sample size formulas for detecting individual, spillover, and overall effects. 
We investigate the roles of the intra-class correlation coefficient
and the probability of treatment allocation on the required number of egocentric networks with a fixed number of network members for each egocentric network and vice-versa.
\end{abstract}

%  Please place your key words in alphabetical order, separated
%  by semicolons, with the first letter of the first word capitalized,
%  and a period at the end of the list.
%

\vspace{2mm}

%  As usual, the \maketitle command creates the title and author/affiliations
%  display 

\maketitle

%  If you are using the referee option, a new page, numbered page 1, will
%  start after the summary and keywords.  The page numbers thus count the
%  number of pages of your manuscript in the preferred submission style.
%  Remember, ``Normally, regular papers exceeding 25 pages and Reader Reaction 
%  papers exceeding 12 pages in (the preferred style) will be returned to 
%  the authors without review. The page limit includes acknowledgements, 
%  references, and appendices, but not tables and figures. The page count does 
%  not include the title page and abstract. A maximum of six (6) tables or 
%  figures combined is often required.''

%  You may now place the substance of your manuscript here.  Please use
%  the \section, \subsection, etc commands as described in the user guide.
%  Please use \label and \ref commands to cross-reference sections, equations,
%  tables, figures, etc.
%
%  Please DO NOT attempt to reformat the style of equation numbering!
%  For that matter, please do not attempt to redefine anything!

\pagenumbering{arabic}
\section{Introduction}
\label{intro}

In the causal inference literature, 
%the individual treatment effect, which quantifies the impact of an intervention or treatment on those who directly receive it, 
the estimation of the treatment effect is well studied under the 
no-interference assumption, 
%stable unit treatment value assumption (SUTVA) \citep{Rubin1974,ImbensRubin2015}. However, an important component of the SUTVA assumption, also known as the no-interference assumption, 
which states that the treatment of one unit cannot affect the outcome of other units \citep{Cox1958, Rosenbaum2007, Rubin1974}. This assumption may be violated when individuals are connected to others through social or physical interactions. For instance, in infectious diseases \citep[P. 211]{Ross1916}, the risk of infection for one person depends not only on their own vaccination status, but also on the vaccination coverage in the population and in particular among more immediate contacts. In education, students enrolled in tutoring programs may affect the school achievement of other students in the same class due to information sharing and peer influence on academic motivation and engagement \citep{Rosenbaum2007}. 

Under interference, 
%untreated individuals are affected by the treatment of others. This effect is commonly called the spillover effect.
%If this is present, 
comparing treated and untreated individuals could be a biased estimation of the treatment effect. By accounting for interference, we can unbiasedly estimate the average effect of receiving the treatment as well as the spillover effect of being exposed to the treated of other units. Disentangling spillover effects from individual treatment effects will provide a more comprehensive understanding of the intervention.
%including how social interactions can amplify the overall impact in the population. 

Research on causal inference methods under interference has been growing in the past two decades, and 
several methods have been developed to assess causal effects 
%individual treatment and spillover effects, also known as direct and indirect effects, 
% The effects of interference have been studied, as the target of inference or as a nuisance, or 
in both randomized experiments and observational studies affected by interference. 
A large body of literature in this field has relied on
the partial interference assumption, which allows interference between individuals within the same group but not across groups (e.g., households, villages, schools) \citep{HudgensHalloran2008,TchetgenVanderWeele2012,Sinclair2012,PerezHeydrich2014,LiuHudgens2014,HalloranHudgens2016,Liu2016}. In recent years, this assumption has been relaxed to more explicitly take into account a more complex form of interference that takes place on a network 
%among units sharing connections 
\citep{SofryginLaan2016,Ogburn2017,AronowSamii2017,Loh2018,Forastiere2020,TchetgenTchetgen2020,Leung2020,Savje2021,Lee2022}. %\citealt{Tchetgen2020};
Under network interference, the potential outcomes of one unit are affected by their own treatment as well as by the treatment received by other individuals directly or, potentially, indirectly connected to them. For example, 
%treatments or preventive measures for infectious diseases are likely to have spillover effects across contact networks by reducing the risk of infection acquisition and onward transmission. 
in behavioral interventions implemented to prompt healthy behaviors, those changing their behaviors as an effect of the received intervention are likely to influence their social ties to do the same \citep{Buchanan2018}. 
%In HIV prevention, educating on people who use drugs about risky and preventive behaviors could affect the drug injection behavior of their friends or drug-sharing partners through behavior modeling, information, or social norms,even if they are not directly exposed to the study intervention \citep{Buchanan2018}.

In this work, we examine an egocentric network-based randomized (ENR) design, where 
two types of study participants are recruited: index participants and their social network members
%, who are nominated by each index participant as people belonging to their personal networks 
(e.g., sex partners, drug partners, those providing social support). The set of network members for each index participant is called their \textit{egocentric network}. The index participants are randomly assigned to the intervention, while their network members are not directly treated but may be exposed to the intervention received by their index participant. For instance,
HIV peer education interventions are designed to leverage a mechanism of peer influence by training index participants on HIV risk reduction and communication skills and encouraging them to disseminate risk reduction information to their sexual/injection network members
\citep{Latkin2009,STEP,CHAT,Buchanan2018}.
%the HIV Prevention Trials Network 037 (HPTN 037) study evaluating a peer education intervention \citep{Latkin2009}, focused on training index participants on HIV risk reduction and communication skills to disseminate risk reduction information to their sexual/injection network members.
An ENR design is often used to evaluate such interventions.
%MOTIVATION
%Egocentric network-based interventions are typically designed to leverage a mechanism of peer influence that is commonly seen and expected in behavioral interventions \citep{Latkin2009,STEP,CHAT,Buchanan2018}. Index participants are trained in disease prevention and promotion of healthy behaviors and are encouraged to train their network members so as to enhance peer influence. 
The assessment not only of the effect of the intervention on index participants but also of the effect of such training and encouragement received by the index participants on the behavioral and infectious disease outcomes of their network members is crucial to fully investigate the impact of such network-based interventions. 
%Randomized experiments have been implemented to evaluate such effects and they imply the randomization of index participants to the intervention. 
However, a formal definition of the causal effects of interest and the identifying assumptions required is needed to be able to estimate such effects. 

In addition, researchers in this field are in need of sample size and power calculations to be able to appropriately design such studies.
The sample size requirements for testing treatment effects in randomized control trials (RCTs) and cluster randomized trials (CRTs) have been well studied \citep{Raudenbush1997,Murray1998,DonnerKlar2000,Wittes2002,HayesMoulton2009,Hemming2017,Walters2019}. 
Meanwhile, \citet{Baird2018} was the first to develop sample size formulas for causal effects under interference. In particular, they developed an optimal design for two-stage (or saturation) designs under a superpopulation framework
%, where in the first stage clusters are assigned to different saturations (proportion of treated individuals), and in the second stage individuals are randomly assigned to treatment according to the saturation assigned to their cluster in the first stage 
\citep{HudgensHalloran2008}. %\citet{Baird2018} provided sample size formulas for the minimum number of clusters required to detect average individual, spillover, and total effects. 
Later, \citet{JiangImai2020} 
developed a
%established a general methodological framework for statistical inference and 
power analysis for the same design under a randomization-based framework.
%, where they provided the unbiased estimators of the causal quantities and their conservative variance estimators.
In this work, we focus instead on the ENR design, for which
%rather than randomized saturation designs or two-stage designs. 
%For this design, as far as we are concerned 
no method for sample size calculation is available.

%\citet{Buchanan2018} formalized the problem of interference in ENR designs under a partial interference assumption and developed a generalized estimating equations method to estimate their causal effects with an interest in the role of the index status. 
Here, 
we make simplifying assumptions of non-overlapping egonetworks and neighborhood interference, i.e., spillover effects are limited to network neighbors. 
Under these assumptions, 
%individuals are either treated and not exposed to the treatment of others, or untreated and exposed to the treated of one of their social connections. 
%This makes the identification and estimation of causal effects under interference easier than other randomized designs. 
%In particular, in this simplified setting, 
we can assess three types of causal effects: 1) the treatment effect
%, defined as the effect of directly receiving the treatment, 
of directly receiving the treatment 2) the spillover effect of being connected to a treated individual;
%, defined as the effect of being connected to a treated individual versus not while being untreated; 
3) the overall effect
%, defined as the effect for all treated and untreated individuals 
of being in an egonetwork where the index participant is treated. 
%versus being in an egonetwork where the index participant is not treated. 
%The definition of the latter effect is specific to the egocentric network-based design, whereas the individual and spillover effects are defined more generally under a neighborhood interference assumption.
We start by developing simple regression-based methods to estimate the individual, spillover, and overall effects in an egocentric network-based randomized design.  We then derive sample size formulas to power studies to detect causal individual, spillover, and overall effects. In particular,
we provide a procedure for calculating the required number of egonetworks with a fixed average number of network members as well as the minimum number of network members for a fixed number of egonetworks.
We consider a study design aimed at testing hypotheses about a single effect and multiple effects using the joint test and the conjunctive test, including individual, spillover, and overall effects, accounting for within-network correlations \citep{Brookes2004,Shieh2009}.

In a closely related article, \citet{Buchanan2018} developed a generalized estimating equations (GEE) method for ENR experiments, relying on a partial interference assumption commonly used in two-stage designs, instead of our neighborhood interference assumption.
%, and defined different causal estimands allowing an effect of being selected as an index participant in addition to that of the treatment. 
In addition, under the partial interference assumption, \citet{Buchanan2018} defined different causal estimands, allowing an effect of being selected as an index participant in addition to that of the treatment.
While their goal was to show how the partial interference assumption can be extended to ENR designs, with a different interpretation of the common causal effects defined in two-stage designs, and to develop GEE estimators, our aim is to derive sample size and power formulas for simple and interpretable causal effects of interest in ENR settings.

%We apply our method to the HIV Prevention Trials Network 037 (HPTN 037) study \citep{Latkin2009}, a network-based study designed to assess an HIV peer education intervention. We consider the study as a pilot data for informing a new study aiming to estimate the individual, spillover, and overall effects.In addition, we show how the sample size would change under slight variations of the assumed parameters.

The remainder of this article is organized as follows. 
%In Section \ref{hptn}, we introduce our motivational study, HPTN 037. 
In Section \ref{egodesign}, we introduce the notation for the egonetwork-based randomized design. The casual estimands and their identification based on the observed data are derived in Section \ref{infer-causal}. In Section \ref{Regression}, we propose the regression-based estimators of the individual, spillover, and overall effects. %and the heterogeneity of the spillover effect. 
%We also derive the sample size formulas for the minimum number of index participants and the network members required for detecting the effects of interest. 
In Section \ref{data}, as an illustrative example, we use the the HIV Prevention Trials Network 037 (HPTN 037) study \citep{Latkin2009}, 
as a pilot study to design a new ENR trial powered to estimate the causal effects of interest.
%a network-based study designed to assess an HIV peer education intervention. We consider the study as a pilot data for informing a new study aiming to estimate the individual, spillover, and overall effects.In addition, we show how the sample size would change under slight variations of the assumed parameters.
%we apply these formulas to the HPTN 037 data. 
Finally, we discuss our findings and potential future work in Section \ref{con}.

\vspace{-0.1cm}

\section{Notation and Egocentric Network-based Randomized Design}
\label{egodesign}

\subsection{Notation}
\label{sec:notation}
%Let us denote by $\mathcal{M}$ a population of interest and by $\mathcal{G}$ an undirected network being the graphical representation of this population, which consists of the pair $(\mathcal{M}, E)$. $E=\{e_{mq}\}_{m,q\in \mathcal{M}}$ is a set of edges representing the links between two units in $\mathcal{M}$. 
%Given a graph $\mathcal{G}$, we denote by$\mathcal{N}_m=\{q\in \mathcal{M}: e_{mq}\in E\}$ the set of units sharing a link with unit $m$, i.e., the `network neighbors' of unit $m$. }

%The networks of the population are constructed by the connected individuals.
In an egocentric network-based randomized study,
a set of index participants %$\mathcal{R}\subset\mathcal{M}$ 
is sampled from the population, denoted by $\mathcal{M}$, and randomly assigned to an intervention.
%with probability $p$, with $0<p<1$. 
In addition, these index participants are asked to provide a list of individuals that they consider to be members of their social network (e.g., friends, sexual partners, drug use partners).
%, i.e., $\mathcal{N}_m$, with $m\in \mathcal{R}$. 
%We denote by $\mathcal{N}$ of size $N=|\mathcal{N}|$ the study population, which consists of the set of index participants and their network members, i.e., 
%$\mathcal{N}=(\mathcal{R},\bigcup_{m\in \mathcal{R}} \mathcal{N}_{m})$.
Network members are not directly given the intervention. Information about baseline characteristics and the outcome of interest is collected for both index participants and network members as part of the baseline and follow-up surveys.
For consistency with the network literature \citep{egonetwork}, we call %`egos' the index participants and `alters' the network members.  
%The set of alters of an ego is commonly referred to as `egocentric network', and here we call 
`egocentric network' the set of network members of each index participant, and `(ego)network' the set of network members of an index participant together with the index participant itself.
In addition, we call intervention (ego)networks the egonetworks with treated egos, and control (ego)networks the egonetworks where the ego is not treated.

\begin{figure}
\centerline{%
\includegraphics[width=\linewidth]{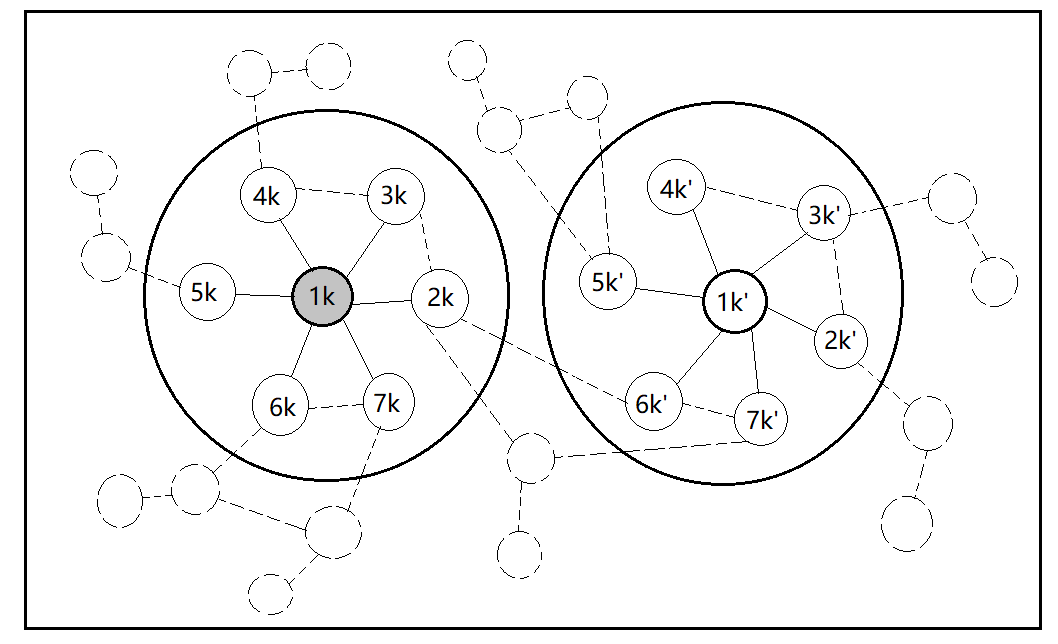}}
%% to include a figure, or to leave a blank space
\caption{
%Double Index for egonetwork notation: all the units are in $\mathcal{M}$; the units in grey circle are in the study population $\mathcal{N}$. Index participants are denoted as $1k$, and network memebers are denoted as $ik$ for $i \neq 1$.
 Egocentric network-based design. The figure represents two egocentric networks denoted by $k$ and $k'$. Index participants are denoted as $1k$ and $1k'$, and their network members are denoted as $ik$ and $ik'$, with $i > 1$. The treated index participant is represented by a grey solid circle. Solid circles represent in-sample units, while dashed circles represent out-of-sample units.  Solid lines are the observed network connections between index participants and their network members, while dashed lines represent network connections that are not observed.
}
\label{fig:dbindex}
\end{figure}

%For the sake of simplicity, we relabel the units in the sample $\mathcal{N}$ relying on a notation commonly used for clustered data.
We denote by $k=1,...,K$ the egonetwork indicator and by $ik$ the individual $i$ in egonetwork $k$, with $i=1, \ldots, n_k+1$. 
For the sake of simplicity,, we let $1k$ represent the $k$th index participant and $ik$, with $i=2,..., n_k+1$, represents a network member of the egonetwork $k$ (in the Appendix \ref{app:notation_remarks}, we clarify the connection between the population notation and this egonetwork notation).
 Under this egonetwork notation, we denote by $\mathcal{N}^*=\{ik\}_{k=1, \dots, K; \,i=1, \ldots, n_k}$ our sample of units, and by $\mathcal{N}^*_k=\{ik \in \mathcal{N}\}_{i=1, \dots, n_k}$ the subsample within each egonetwork $k$.
Finally, let $R_{ik}$ be and indicator for whether unit $ik$ is an index participant ($R_{ik}=1$) or a network member ($R_{ik}=0)$. Given our egocentric notation, it follows that $R_{1k}=1$ and $R_{ik}=0$ for all $i>1$. 
In Figure \ref{fig:dbindex}, we provide a graphical representation of the ENR design. 
%units in the grey circle are the participants in $\mathcal{N}$, and they are labelled using $ik$ notation. 

Let us now define the variables of interest. Let $Y_{ik}$ denote the outcome variable for individual $i$ in egonetwork $k$. 
Regardless of the treatment assignment mechanism, we denote by
$Z_{ik}$ the treatment variable so that $Z_{ik} =1$ when individual $ik$ is treated and $Z_{ik}=0$ otherwise. 
In an egocentric network-based randomization design, only index participants can be treated while network members cannot, i.e.,
$Z_{1k}\in \{0,1\}$ and $Z_{ik}=0$ for $i>1$. 
% Note that all units that are not in the study sample are considered as not treated because they cannot receive the intervention.
We assume that index participants are randomly assigned to the intervention with probability $p$, with $0<p<1$, following a Bernoulli randomization, i.e., $Pr(Z_{ik}=1|R_{ik}=1)=p$. On the contrary, network members cannot receive the intervention, i.e., $Pr(Z_{ik}=1|R_{ik}=0)=0$.

\subsection{Non-overlapping egonetworks}
\label{sec:nonoverlap}
%With a slight abuse of notation, 
We denote by $\mathcal{N}_{ik}$ the network neighborhood of unit $i$ in egonetwork $k$. 
In an ENR design, in the sample $\mathcal{N}$ we only observe information on connections of index participants, i.e., $\mathcal{N}_{1k}$,
%\footnote{$\mathcal{N}_{1k}$ coincides with $\mathcal{N}_{r(k)}$ using the population indexing.}, 
whereas connections of network members $ik$, with $i=2, \ldots, n_k$
%, i.e.,  $\mathcal{N}_{n(ik)}$, 
are in general not observed. 
%Under the assumption that the network $\mathcal{G}$ is undirected,  
In an undirected network, 
the only connection that we observe for network members is the one with their index participants. However, in addition to possible unobserved connections with out-of-sample individuals, in principle network members $ik$ can be linked to other index participants $1k'$, with $k'\neq k$. 
Furthermore, connections among network members in the same egonetwork are not observed.
Although by network transitivity it is likely that the peers of an index participant are also connected to each other, a fully connected egonetwork is not guaranteed and some pairs of network members of the same egonetwork may not be linked.

We make here a simplifying assumption that will be needed 
%for the definition of the interference mechanism and 
for the identification of causal effects.
We assume that network members are only connected to one index participant in the sample and that index participants are not connected among themselves. 
Formally, we denote by $\mathcal{N}^*_{ik}\subset \mathcal{N}^*$ the network neighborhood of unit $i$ in egonetwork $k$ only including in-sample units.
Then, we make the following assumption.
\begin{assumption}[Non-overlapping Egonetworks]
\label{ass:nonoverlap}
$\mathcal{N}^*_{ik}\subset \mathcal{N}^*_{k} \quad \forall ik \in \mathcal{N}^*$
\end{assumption}

\noindent The plausibility of this assumption in HPTN 037 has already been discussed in \cite{Buchanan2018}. %, although the assumption was not formalized.
%and other egocentric designs. 
The non-overlapping egonetworks assumption can be guaranteed by design by selecting index participants whose egocentric networks are unlikely to overlap.

%Given the graph $\mathcal{G}$ and the treatment vector $\Z$ in the sample,
We let 
$G_{ik}$ be the number of treated network neighbors for unit $i$ in egonetwork $k$. 
%i.e., $G_{ik}=\sum_{q \in \mathcal{N}_{h(ik)}} Z_{q}= 1 $ 
As already mentioned, network members may be connected with other individuals not in the sample. However, because we assume that individuals that are not in the sample cannot receive the treatment,
%Note that all units that are not in the sample are considered as not treated, that is, $Z_{m}=0\, \forall m\notin \mathcal{N}$. 
$G_{ik}$ is given by the number of treated network neighbors in the sample, i.e.,  $G_{ik}=\sum_{\ell k \in \mathcal{N}^*_{ik}} Z_{\ell k} $.
Furthermore, thanks to the non-overlapping network assumption (Assumption \ref{ass:nonoverlap} and that in the egocentric network-based design network members cannot be treated), we have 
that the number of treated network neighbors is equal to the number of treated individuals in the same egonetwork, excluding the individual itself, i.e., $G_{ik}=\sum_{\ell k \in \mathcal{N}^*_k, \ell\neq i} Z_{\ell k}= 
(1-R_{ik})Z_{1k}$.
% With this notation, we have that $Y_{ik}=Y_{h(ik)}$, $\mathbf{X}_{ik}=\mathbf{X}_{h(ik)}$, $Z_{ik}=Z_{h(ik)}$, and $G_{ik}=G_{h(ik)}$.
As a consequence, under the ENR design and under Assumption \ref{ass:nonoverlap}, index participants $1k$ have $Z_{1k}=\{0,1\}$ and $G_{1k}=0$, because they can be treated but cannot have any treated network neighbor, while network members $ik$, with $i>1$, have $Z_{ik}=0$ and $G_{ik}=Z_{1k}=\{0,1\}$, with the value of $G_{ik}$ depending on whether they are in an intervention or control network.

\vspace{-0.5cm}

\section{Interference and Causal Estimands}
\label{infer-causal}

\subsection{Neighborhood Interference}
Under the potential outcome framework, we denote by $Y_{ik}(\Z)$ the potential outcome of individual $i$ in network $k$ under the sample treatment vector $\Z$. 
%in the sample of dimension $N\times 1$. 
Here, we relax the no-interference assumption, allowing for the outcome of an individual $i$ in egonetwork $k$ to be affected by their own treatment $Z_{ik}$ and also by the number of treated network neighbors $G_{ik}$.  
%that is, interference is restricted to the network neighborhood. 
In the causal inference literature, this assumption is known as `neighborhood interference', which restricts interference to the network neighborhood. In addition, we are assuming a unit's potential outcome depends on a specific function of the neighbors' treatment %interference occurs through a specific function of the   with a particular type of exposure mapping function being the number of treated neighbors 
\citep{AronowSamii2017,Ogburn2017,Forastiere2020}.
Formally:
\begin{assumption}[Neighborhood Interference]
\label{ass:neighint}
Given $\Z$ and $\Z'$ such that $Z_{ik}=Z'_{ik}$ and $G_{ik}=G'_{ik}$, then
$Y_{ik}(\Z)=Y_{ik}(\Z')$.
\end{assumption}
\noindent Under this assumption, we can index potential outcomes only by $Z_{ik}$ and $G_{ik}$: $Y_{ik}(z,g)$ denotes the potential outcome of participant $i$ in network $k$ under individual treatment $Z_{ik}=z$ and number of treated neighbors $G_{ik}=g$. 
%\textcolor{blue}{Note that under the egonetwork-based design, for treated index participants, we have $Z_{1k}=1$ and $G_{1k}=0$; for network members in intervention network, we have $Z_{ik}=0$ and $G_{ik}=1$ for any $i>1$; for participants in networks without intervention, we have $Z_{ik}=0$ and $G_{ik}=0$.}
%
Assumption \ref{ass:neighint} 
rules out the possibility that individuals' behaviors are indirectly affected by the intervention received by other individuals with whom they are not directly connected 
%but may be indirectly connected through their peers or through higher-order connections 
(e.g., friends of friends). This assumption is satisfied if behavioral influence takes time to travel through the network and the time when the behavioral outcome is measured in the study only allows for influence to occur between network neighbors. In egonetwork studies, Assumption  \ref{ass:neighint} is also plausible if egonetworks are sufficiently distant in the network, such that even if interference could occur  beyond the network neighbors, the effect of the treatment received by one index participant would not reach individuals in other egonetworks.
In HPTN 037 the plausibility of the neighborhood assumption is supported by both egonetwork distances and the outcome measure.

%%SENT TO THE APPENDIX
% It is worth mentioning the relationship between the neighborhood interference assumption (Assumption \ref{ass:neighint}) and the common stratified interference assumption. The latter rules interference between groups allowing interference only between groups (partial interference) and assumes that interference depends only on the number of treated individuals in the same group and not on who they are \citep{HudgensHalloran2008,Buchanan2018}.
% Under the specific egocentric network-based design, where only index participants can be treated, and  
% under the non-overlapping egonetworks assumption, the neighborhood interference assumption (Assumption \ref{ass:neighint}) is equivalent to the stratified interference assumption with groups being the egonetworks. In fact, as already shown, $G_{ik}$, which is by definition equal to the number of network neighbors, i.e., $G_{ik}=\sum_{\ell k \in \mathcal{N}^*_{ik}} Z_{\ell k} $, in this setting will also be equal to the number of the other individuals in the same egonetwork who are treated, i.e., $G_{ik}=\sum_{\ell k \in \mathcal{N}^*_k, \ell\neq i} Z_{\ell k} $.
% In the egocentric network-based study, $Z_{ik}$ and $G_{ik}$ can only take value 0 or 1 since at most one participant can be treated in an egonetwork.
%--------------------------%

\vspace{-0.6cm}

\subsection{Causal Estimands}
\label{Causal-Est}
%Now we can define the causal estimands of interest. 
We define the \textit{average (individual) treatment effect (AIE)} as the average effect of receiving the treatment when network neighbors are all untreated, that is:
\begin{equation}
\label{eq:tau}
   \tau= \E\left[Y_{ik}(1,0)-Y_{ik}(0,0)\right].
\end{equation}
Note that the expectation is taken over the distribution of potential outcomes in the population $\mathcal{M}$, under the common superpopulation perspective to causal inference (\citealt{whatif}).
Similarly, we define the \textit{average spillover effect (ASpE)} as the average effect of having one treated network neighbor versus none while the individual is untreated:
\begin{equation}
\label{eq:delta}
    \delta=\E\left[Y_{ik}(0,1)-Y_{ik}(0,0)\right].
\end{equation}

We are also interested in the \textit{overall effect}, defined as the effect of being in an egonetwork where one unit is treated (intervention egonetwork) versus being in an egonetwork where no one is treated (control egonetwork). We can write the causal estimand as the average difference between the potential outcomes of the treated 
%unit and the 
untreated units in an intervention network and the potential outcomes of the untreated units in the control networks:
\begin{equation}
%\begin{aligned}
\label{eq:overall}
\begin{array}{ll}%
   O =  \E \left[\sum_{z=0}^1\sum_{g=0}^1 Y_{ik}(z,g)Pr(Z_{ik}=z,G_{ik}=g|Z_{1k}=1)\right]  \\ 
  \quad   \quad   - \E \left[\sum_{z=0}^1\sum_{g=0}^1  (Y_{ik}(z,g)Pr(Z_{ik}=z, G_{ik}=g|Z_{1k}=0)\right] \\  
  \quad   = \tau Pr(R_{ik}=1) +\delta Pr(R_{ik}=0),
 %  \end{aligned}
\end{array}
\end{equation}
% \noindent Given Assumptions 1-3, the overall effect can be identified from the observed data as $$O=\E \left[ \frac{1}{n_k+1}\biggl(\E[Y_{ik}| Z_{ik}=1, G_{ik}=0] + n_k \E[Y_{ik}| Z_{ik}=0, G_{ik}=1]\biggr)\right]-\E[Y_{ik}|Z_{ik}=0, G_{ik}=0].$$
The proof of Equation (\ref{eq:overall})
%and of the identification of the overall effect are 
is given in the Appendix \ref{iden-o}.
When the egonetwork size is constant, i.e., $n_k=n \quad \forall k$, the overall effect is equal to $O=\frac{1}{n+1}(\tau +n\,\delta)$.

\vspace{-0.3cm}

 \subsection{Identifying Assumption and Identification of Causal Effects}
 \label{sec:identification}

Given the characteristics of an egocentric network-based design, 
%and 
under Assumptions \ref{ass:nonoverlap} and \ref{ass:neighint}, 
%and \ref{ass:unc}, %%\ref{ass:unc}, and \ref{ass:cons},  
and under additional assumptions detailed in the Appendix \ref{sec:identification_assumptions}, including unconfoundedness, consistency, and random sampling,
we can identify the causal effects of interest. 
In particular, we can identify AIE and ASpE from the observed data as $\tau = \E[Y_{ik}|Z_{ik}=1]-\E[Y_{ik}|Z_{ik}=0, G_{ik}=0]$ and $\delta = \E[Y_{ik}| G_{ik}=1]-\E[Y_{ik}|Z_{ik}=0, G_{ik}=0]$.
%The proofs of identification of the AIE and ASpE are given in S.3.1 and S.3.2 of the Supplementary Materials, respectively. % Section \ref{iden-tau} and \ref{iden-delta}, respectively. %
Similarly, the overall effect can be identified from the observed data as $O= \E[Y_{ik}| Z_{ik}=1]Pr(R_{ik}=1) + \E[Y_{ik}| G_{ik}=1]Pr(R_{ik}=0)-\E[Y_{ik}|Z_{ik}=0, G_{ik}=0].$
%The proof of the identification of the overall effect is given in S.3.3 of the Supplementary Materials.
The proofs of the identification are given in the Appendix \ref{identification}.
% $O = \sum\limits^{K}_{k=1}\frac{\tau+n_k \delta}{(n_k+1)N}$
%since I write it in a sumation of K cluster, so total number of cluster is K

% \begin{assumption}[\textcolor{blue}{Modeling assumptions}]
% \label{ass:model}
% \textcolor{blue}{We assume that $Y_{ik}(z,g)$ follows the normal distribution with mean taking value of $\gamma$ for $Y_{ik}(0,0)$, $\tau$ for $Y_{ik}(1,0)$, and $\delta$ for $Y_{ik}(0,1)$, where $\gamma$ denoted as mean of individuals in networks without intervention. We also assume that $Y_{ik}(z,g)$ has exchangeable correlation structure. }
% \end{assumption}

% add some elaberations? 

%\subsection{\textcolor{blue}{Modeling assumptions}}

\vspace{-0.5cm}
\section{Regression-based Estimators and Sample Size Calculation \label{Regression}}

In this section, we propose regression-based estimators for AIE, ASpE, and $O$. We also derive formulas for the required number of index participants for detecting each causal effect or causal effects with pre-specified power and Type I error rate. To do so, we assume that all index participants have the same number of network members, i.e., $n_k = n$ for $k=1,...,K$.
This assumption can be achieved by design by limiting the number of network members that each index participant can nominate. If $n$ is sufficiently small, it is plausible that index participant would nominate at least $n$ network members. If the list of network members is censored, we can assume that censoring is non-informative given that the treatment is randomized to index participants.

\vspace{-0.1cm}

%\begin{figure}[!p]
%\centering
%\minipage{0.8\textwidth}%
 % \includegraphics[width=\linewidth]{figure/design_1.jpg}
 % \caption{Egonetwork-based design and subsets of data used for the identification of each effect. }
 % \label{fig:design}
%\endminipage
%\end{figure}

\begin{figure}
\centerline{%
\includegraphics[width=\linewidth]{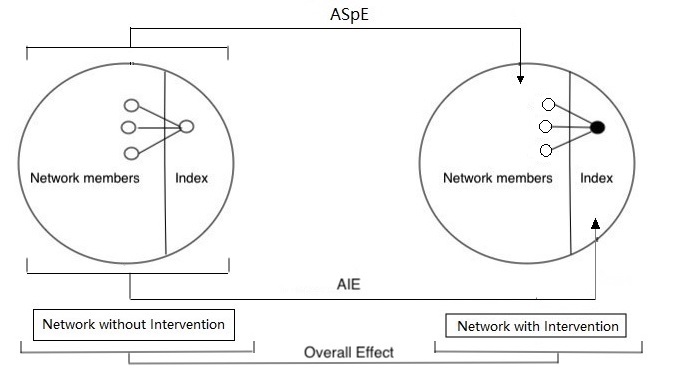}}
%% to include a figure, or to leave a blank space
\caption{Egonetwork-based design and subsets of data used for the identification of each effect.}
\label{fig:design}
\end{figure}

\vspace{-0.2cm}

\subsection{Statistical Models for the Average Treatment, Spillover, and Overall Effects}

%\subsubsection{Statistical Model}

We introduce the regression model for estimating the average individual and spillover effects, $\tau$ and $\delta$, based on the identification results in Section \ref{sec:identification}.
Under the stated assumptions, the AIE can be identified by comparing the outcomes of index participants in the intervention egonetworks (with $Z_{ik}=1$) to the outcomes of all individuals in the control egonetworks (with $Z_{ik}=0$ and $G_{ik}=0$), while the ASpE can be estimated by comparing the outcomes of networks members in the intervention egonetworks (with $G_{ik}=1$) to the outcomes of all individuals in the control egonetworks (with $Z_{ik}=0$ and $G_{ik}=0$) (Figure~\ref{fig:design}). Furthermore, we can identify the overall effect by comparing the outcomes of both the index participants and their network members in the intervention egonetworks to the outcomes of both the index participants and their network members in the control egonetworks. Alternatively, once $\tau$ and $\delta$ have been estimated, the overall effect can be estimated by  $\widehat{O}=\frac{1}{(n+1)}(\widehat{\tau} +n \widehat{\delta})$.

For $k=1,...,K$ and $i =1,...,n+1$, we let
\begin{equation}
Y_{ik}=\gamma + \tau Z_{ik} + \delta G_{ik} + u_k + \epsilon_{ik}, \label{model-1}
\end{equation}
where we assume that the residual error $\epsilon_{ik}\sim N(0,\sigma^2_e)$ and the random egonetwork effect $u_k \sim N(0,\sigma^2_u)$.
%have zero means  and variance $\sigma^2_e$ and $\sigma^2_u$, respectively.
 In Model \ref{model-1}, the coefficient $\gamma$ is the mean outcome of individuals in networks without intervention, i.e., $\gamma=\E (Y_{ik}|Z_{ik}=0,G_{ik}=0)$.  The coefficient  $\tau$  is equal to $\E[Y_{ik}|Z_{ik}=1, G_{ik}=0]-\E[Y_{ik}|Z_{ik}=0, G_{ik}=0]$, which, under the previously stated assumptions, identifies the causal effect $\tau$ in Equation \eqref{eq:tau}, i.e.,  the AIE. Finally, the coefficient $\delta$ is equal to $\E[Y_{ik}|Z_{ik}=0, G_{ik}=1]-\E[Y_{ik}|Z_{ik}=0, G_{ik}=0]$, which, in turn, under the previously stated assumptions, identifies the ASpE denoted as $\delta$ and defined in Equation (\ref{eq:delta}).
 
To proceed, we let $\btheta = (\gamma,\tau,\delta)'$ and  further define  $\Y_k = (Y_{1k},...,Y_{ik},...,Y_{(n+1)k})'$ and $\D_{k} = (1,\Z_{k},\mbG_{k})$, where $\Z_{k} = (Z_{1k},...,Z_{(n+1)k})'$ and $\mbG_{k} = (G_{1k},...,G_{(n+1)k})'$. By design, $\D_{k}$ takes the specific forms: when $Z_{1k}=1$, $G_{1k}=0$ and $G_{ik} =1$ for $i >1$; when $Z_{1k }= 0$, $Z_{ik}$ and $G_{ik}$ are always 0.
By the sampling mechanism, as well as in Model \ref{model-1}, which in fact, models the distribution of the observed outcomes given the sampling mechanism and randomization, the observed outcomes of two individuals in different egonetworks are independent.  The covariance between the observed outcomes of two members in the same egonetwork is $\Cov(Y_{ik},Y_{i'k}|\Z_{k},\mbG_{k}) = \sigma^2_u $ for $i \neq i'$ given $\Z_{k}$ and $\mbG_{k}$. Finally, the total variance of $Y_{ik}$, denoted by $\sigma^2_{Y}$, is equal to $\sigma^2_u +\sigma^2_e$.
Then, the intra-class correlation (ICC) between $Y_{ik}$ and $Y_{i'k}$, for $i \neq i'$, conditional on $\Z_{k}$ and $\mbG_{k}$ is
$
\rho_{Y} = \frac{\sigma^2_u}{\sigma^2_u+\sigma^2_e}.
$
As a result, the variance of the outcome vector $\Y_{k}$ for the $k$th egonetwork is $\Var(\Y_k|\Z_{k},\mbG_{k})=\sigma^2_{Y}\cdot V_k$ with $V_k = (1-\rho_{Y})I_{(n+1)}+ \rho_{Y} J_{(n+1)}$, where $I_{(n+1)}$ is a $(n+1) \times (n+1)$ identity matrix, $J_{(n+1)}$ is a $(n+1) \times (n+1)$ matrix where all elements are 1. Under this variance-covariance structure, we can estimate the parameter vector $\btheta$ of Model \eqref{model-1} using generalized least squares (GLS) as
$%\[
\hat{\btheta} =  \left(\sum\limits^{K}_{k=1}\D^{'}_{k} V^{-1}_k \D_{k}\right)^{-1} \left(\sum\limits^{K}_{k=1} \D^{'}_{k} \Y_{k} \right)
$ %\]
with
$
\Var(\hat{\btheta}) = \sigma^2_{Y} \U^{-1}_{Ik},
$
where $\U_{Ik}=\sum\limits^K_{k=1} \D^{'}_{k} V^{-1}_k \D_{k}$.  
Under regularity conditions, as $K \to \infty$ and $n$ is fixed, $\sqrt{K}(\hat{\btheta}-\btheta)$ is asymptotically normally distributed as $N(0, \Sigma_I)$, where $\Sigma_I = \lim_{K \to \infty} \sigma^2_{Y} (\U_{Ik}/K )^{-1}=\sigma^2_{Y} \U_I^{-1}$ with $\U_I = \lim_{K \to \infty} \frac{1}{K} \U_{Ik} $. 

 We now use $\Sigma_{I}$ to construct the formulas for finding the required number of index participants for detecting different causal effects.
In Appendix \ref{Var-X}, % \ref{Var-X},
we show that 
\[
\U_I = \left[\begin{array}{c|cc}
 \{c+(n+1)d\} (n+1)  &  \{c+(n+1)d\} p & \{c+(n+1)d\}  n p \\
 \hline
 \{c+(n+1)d\} p & (c+d) p & n p d \\
 \{c+(n+1)d\}  n p  & n p d & (c+nd) n p \\
\end{array} \right]
\]
with $c  = \frac{1}{1-\rho_{Y}}$ and $d = -\frac{\rho_{Y}}{(1-\rho_{Y})(1+n\rho_{Y})}$. Then, the resulting lower-right block of $\Sigma_I$ is the covariance corresponding to $\tau$ and $\delta$, denoted as $\Sigma_{\tau\delta}$
\[
\Sigma_{\tau\delta} =\sigma^2_{Y} \begin{bmatrix}
\frac{c m_2 +nd\sigma^2_Z}{cp\sigma^2_Z\{c+d(1+n)\}} &  \frac{c(p-m_1) - d\sigma^2_Z}{cp\sigma^2_Z\{c+d(1+n)\}}\\
\frac{c(p-m_1) - d\sigma^2_Z}{cp\sigma^2_Z\{c+d(1+n)\}} & \frac{cm_1+d\sigma^2_Z}{ncp\sigma^2_Z\{c+d(1+n)\}} \\
\end{bmatrix},
\]
where $\sigma^2_Z = p(1-p)$, $m_1 = p(1-\frac{p}{n+1})$ and $m_2 =p(1-\frac{np}{n+1})$. The derivation of $\Sigma_{\tau\delta}$ can be found in the Appendix \ref{Sig-1}.

\subsection{Calculation of the Minimum Number of Networks \label{Cal_K}}

Based on Model (\ref{model-1}) and the GLS estimator, we consider procedures for testing several hypotheses of potential interest related to the AIE, ASpE, and $O$.
 We then derive the required number of index participants to ensure adequate power to test the hypotheses of interest, given prespecified Type I error rate, significance level, and effect sizes.
In the Appendix \ref{nMDE}, we further provide formulas for the required number of network members given a specified number of index participants, a hypothesized effect size, a desired power and Type I error, as well as the minimum detectable effect size (MDE) given a specified number of index participants and network members, a desired power and Type I error.

%Hypothesis Test 1: 
\textit{The AIE hypothesis test (HIE).} Here we test the hypothesis of no AIE; that is, $H_0: \tau=0$, against the alternative hypothesis $H_1: \tau \neq 0$. To test this hypothesis, we use the two-sided Z-test statistic: $T_{\tau} =\sqrt{K}(\hat{\tau}/\hat{\sigma}_{\tau})$, where $\sigma^2_{\tau}$ is the asymptotic variance of $\hat{\tau}$ from $\Sigma_{\tau\delta}$ (the top-left element in $\Sigma_{\tau\delta}$), 
$
    \sigma^2_{\tau} = \frac{\sigma^2_{Y}\{n(1-p)(1-\rho_{Y})+(1+n\rho_{Y}) \}}{(n+1)\sigma^2_Z}
$.

Given the GLS estimator for $\btheta$ in Model \eqref{model-1}, $T_{\tau}$ asymptotically follows a standard normal distribution under the null hypothesis. 
Assume that the effect size of $\tau$ is $\Delta_{\tau}$. Given a Type I error rate $\alpha$, the probability of rejecting $H_0$ when it is true is $P(|T_{\tau}|>z_{1-\alpha/2} \vert \tau = 0)$ with critical value $z_{1-\alpha/2}$. Then, the power of the test, 
$ \pi_{\tau}=P\{|\sqrt{K}(\hat{\tau}-\Delta_{\tau})/\hat{\sigma}_{\tau})|\geq  z_{1-\alpha/2} -\sqrt{K} \Delta_{\tau}/\sigma_{\tau}\vert \tau = \Delta_{\tau}\} =1- \Phi(z_{1-\alpha/2}-\sqrt{K}\Delta_{\tau}/\sigma_{\tau})+\Phi(z_{\alpha/2}-\sqrt{K}\Delta_{\tau}/\sigma_{\tau})$, 
where $\Phi(\cdot)$ is the cumulative distribution function of the standard normal distribution and $z_{\pi} = \Phi^{-1}(\pi)$ for any power $\pi  \in [0,1]$. When the sample size of the two-sided test is calculated, the minuscule region associated with one of the tails based on whether the effect is positive or negative is often ignored. Thus,
given the required power and Type I error rate $\alpha$, we solve $\pi_{\tau}$ ignoring the minuscule region for $K$ to obtain the required number of index participants for a specified effect size $\Delta_{\tau}$ for the HIE, that is,

%$\pi_1-\pi_2 = 0.8$

%$z_{\pi_1} = z_{1-\alpha/2}-\sqrt{K}\Delta_{\tau}/\sigma_{\tau}$

%$z_{\pi_2} = z_{\alpha/2}-\sqrt{K}\Delta_{\tau}/\sigma_{\tau}$

\begin{equation}
\begin{array}{cl}
 K_{{\tau}} &=\frac{\sigma^2_{\tau}(z_{1-\alpha/2}+z_{\pi})^2}{\Delta_{\tau}^2}
 %=  \frac{\sigma^2_{Y}\{n(1-p)(1-\rho_{Y})+(1+n\rho_{Y}) \} (z_{1-\alpha/2}+z_{\pi})^2}{(n+1)\sigma^2_Z \Delta_{\tau}^2} \\
=  \frac{\sigma^2_{Y}[n\{p (\rho_Y -1)+1\}+1] (z_{1-\alpha/2}+z_{\pi})^2}{(n+1)\sigma^2_Z \Delta_{\tau}^2}. 
\end{array}
\label{K-tau}
\end{equation}

\textit{The ASpE hypothesis test (HSpE).} Here, we focus on the hypothesis of no ASpE, $H_0: \delta=0$, against the alternative hypothesis $H_A: \delta\neq 0$. To test this hypothesis, we use the two-sized Z-test statistic: $T_{\delta} = \sqrt{K}(\hat{\delta}/\hat{\sigma}_{\delta})$, where
$
\sigma^2_{\delta}
 = \frac{\sigma^2_{Y}\{(1-p)(1-\rho_{Y})+n(1+n\rho_{Y})\}}{n(n+1)\sigma^2_Z}, 
$
is the bottom-right element of $\Sigma_{\tau\delta}$ corresponding to $\hat{\delta}$. 

Similar to above, $T_{\delta}$ follows a standard normal distribution under $H_0$. For a specified effect size $\Delta_{\delta}$ for the ASpE, a given Type I error rate $\alpha$, the power of the test is $\pi_{\delta} =1- \Phi(z_{1-\alpha/2}-\sqrt{K}\Delta_{\delta}/\sigma_{\delta})+\Phi(z_{\alpha/2}-\sqrt{K}\Delta_{\delta}/\sigma_{\delta})$. We can solve this equation for $K$ to obtain the required number of networks to achieve adequate power for the given Type I error rate, $\alpha$, and ignoring the minuscule region, by,
\begin{equation}
K_{{\delta}}=\frac{\sigma^2_{\delta}(z_{1-\alpha/2}+z_{\pi})^2}{\Delta_{\delta}^2} =  \frac{\sigma^2_{Y}\{(1-p)(1-\rho_{Y})+n(1+n\rho_{Y}) \} (z_{1-\alpha/2}+z_{\pi})^2}{n(1+n)\sigma^2_Z \Delta_{\delta}^2}. \label{K-delta}
\end{equation}

\textit{The AIE and ASpE joint hypothesis test (HISpJ).} When we are interested in testing both individual and spillover effects simultaneously, as would usually be the case in an egocentric network-based randomized design, we can construct a 
%two-sized 
two degree of freedom Wald test for the joint hypothesis $H_0: \btheta_J  = 0$, where $\btheta_J = (\tau,\delta)'$.
The Wald test statistic of HISpJ is 
$
Q_J = K\hat{\btheta}'_J \hat{\Sigma}_{\delta \tau}^{-1}\hat{\btheta}_J
$.
From what has previously been shown about the asymptotic distribution of $\hat{\btheta}$, it follows that $\hat{\btheta}_J$ has a multivariate normal distribution asymptotically, with mean $\btheta_J$ and covariance $\Sigma_{\tau\delta}$. Then, $Q_J$ is asymptotical approximately $\chi^2$ distributed. Given a Type I error $\alpha$ and effect size $\Delta_J$, the power of this test is
$\pi_{J} =P\{Q_J \geq \chi^2_{1-\alpha}(2) |\btheta_J = \Delta_J  \} $. Then, the required number of index participants for HISpJ
 to have power $\pi_{J}$ at Type I error rate $\alpha$ can be obtained by solving $\pi_{J} \geq \pi$ for $K$.
In the Appendix \ref{N-HISpJ}, 
we show that the required number of index participants is
$
K \geq \frac{\upsilon(\chi^2_{1-\alpha}(2),\pi,2)}{\Delta_J'\Sigma_{\delta \tau}^{-1}\Delta_J},
$
where $\upsilon(q,\pi,p_T)$ is the non-centrality parameter of the non-central $\chi^2$ distribution with two degree of freedom whose $1-\pi$ quantile is equal to $q$. Then, the resulting required number of index participants for HISpJ is
\begin{equation}
K_{J} =\frac{\upsilon(\chi^2_{1-\alpha}(2),\pi,2) \sigma^2_{Y}(1+n\rho_Y)}{\sigma^2_{Z}(\Delta_{\tau}^2+n\Delta_{\delta}^2)}.  \label{K-td}
\end{equation}

\textit{The AIE and ASpE conjunctive hypothesis test (HISpC).} The HISpJ rejects the null hypothesis when at least one of AIE and ASpE has an effect on the outcome. 
However, sometimes we are interested in the case that the intervention is effective in terms of both the AIE and ASpE; that is, both causal effects are non-zero.
A conjunctive hypothesis test can be used for this purpose with $H_0: \tau = 0$ or $\delta = 0$ against the alternative hypothesis $H_A: \tau \neq 0$ and   $\delta \neq 0$ \citep{Tian2022}. To test this hypothesis, we use a bivariate test statistic $Q_{C}=(T_{\tau}, T_{\delta})^{T}$, where $T_{\tau}$ and $T_{\delta}$ correspond to the HIE and HSpE, respectively.
Then $Q_{C}$ follows a bivariate normal distribution
$
Q_{C} \overset{d}{\to} N \left ( \begin{bmatrix}
\sqrt{K} \Delta_{\tau}/\sigma_{\tau} \\
\sqrt{K}\Delta_{\delta}/\sigma_{\delta}
\end{bmatrix}, \Omega =  \begin{bmatrix}
1 & \frac{\sigma_{\tau\delta} }{\sigma_{\tau}\sigma_{\delta} }\\
\frac{\sigma_{\tau\delta} }{\sigma_{\tau}\sigma_{\delta} } & 1
\end{bmatrix}
\right),
$
where $\sigma_{\tau\delta}=K Cov(\hat{\tau},\hat{\delta})$ with 
$
Cov(\hat{\tau},\hat{\delta}) = \frac{\sigma^2_Y\{p(1+n\rho_Y)+(1-p)(n+1)\rho_Y \}}{(n+1)\sigma^2_Z}.
$
Then the power formula for the two-sided conjunctive test is
\begin{equation}
\begin{aligned}
\pi_{C}  &= \Pr[\{|T_{\tau}|>Z_{1-\alpha/2} \} \cap \{|T_{\delta}|>Z_{1-\alpha/2} \}  ]\\
& = \int^{\infty}_{Z_{1-\alpha/2}} \int^{\infty}_{Z_{1-\alpha/2}} u(T_{\tau},T_{\delta})dT_{\tau}dT_{\delta} + \int^{\infty}_{Z_{1-\alpha/2}} \int^{Z_{\alpha/2}}_{-\infty} u(T_{\tau},T_{\delta})dT_{\tau}dT_{\delta} \\
&+\int^{Z_{\alpha/2}}_{-\infty}\int^{\infty}_{Z_{1-\alpha/2}}  u(T_{\tau},T_{\delta})dT_{\tau}dT_{\delta} +\int^{Z_{\alpha/2}}_{-\infty}\int^{Z_{\alpha/2}}_{-\infty} u(T_{\tau},T_{\delta})dT_{\tau}dT_{\delta} 
\end{aligned}, \label{K-HISpC}
\end{equation}
where $u(T_{\tau},T_{\delta})$ is the PDF of the bivariate normal distribution of $Q_{C}$ as discussed above. To find the required member of networks ($K_{C}$), we solve \eqref{K-HISpC}. This power formula can be used to numerically calculate the required number of networks because there is no closed form of this formula. A series
of increasing integers $K$ can be plugged into the equation to compute the power after specifying the values of $\rho_Y$, $n$, $p$, $\Delta_{\tau}$, and $\Delta_{\delta}$.

\textit{The overall effect hypothesis test (HOE).} When we are interested in testing the overall effect, we can construct a two-sided Z-test: $H_0: (\tau+n\delta)/(n+1)=0$. To test this hypothesis, we use the Z-test statistic: $T_{o}=\sqrt{K}\{(\hat{\tau}+n\hat{\delta})/(n+1)\}/\hat{\sigma}_{o}$,
where $\sigma^2_{o} = \frac{\sigma^2_Y(1+n\rho_Y)}{(n+1)\sigma^2_Z}$.
%\[
%\begin{aligned}
%\sigma^2_{o} &= \frac{1}{(n+1)^2}\left\{\sigma^{2}_{\tau}+n^2\sigma^{2}_{\delta} +2n\Cov(\hat{\tau},\hat{\delta}) \right\} \\
%& = \frac{\sigma^2_Y(1+n\rho_Y)}{(n+1)\sigma^2_Z}\\
%\end{aligned}.
%\]
%with
%\[
%\Cov(\tau,\delta) = \frac{\sigma^2_Y\{p(1+n\rho_Y)+(1-p)(n+1)\rho_Y \}}{(n+1)\sigma^2_Z}.
%\]
$T_{o}$ is a linear transformation of $\hat{\tau}$ and $\hat{\delta}$, and follows a standard normal distribution when the effect size of the overall effect, $\Delta_{o}=(\Delta_{\tau}+n\Delta_{\delta})/(n+1)$ is zero. Similar to HIE and HSpE, given a Type I error $\alpha$, the power of the test is  $\pi_{o} =1- \Phi(z_{1-\alpha/2}-\sqrt{K}\Delta_{o}/\sigma_{o})$. To find the required member of networks ($K_{o}$) at power $\pi_{o}$, we obtain
\begin{equation}
K_{o}=\frac{\sigma^2_{o}(z_{1-\alpha/2}+z_{\pi})^2}{\Delta_{o}^2} =  \frac{\sigma^2_{Y}(1+n\rho_Y) (z_{1-\alpha/2}+z_{\pi})^2}{(n+1)\sigma^2_Z \Delta_{o}^2}. \label{K-o}
\end{equation}

\subsection{Investigation of the Sample Size for Each Hypothesis Test as a Function of Features of the Egocentric Network-based Design}
%Theoretical properties of features of the egocentric network based design}
%\subsection{\textcolor{blue}{Numerical Analysis}}

To learn more about how $K$ changes with the other parameters, we investigate these relationships through some simulations. From \eqref{K-tau} and \eqref{K-delta}, we observe that $K_{{\tau}}$ and $K_{{\delta}}$ are functions of 
%the probability of index treatment 
$p$, the network size $n$, the intra-class correlation $\rho_Y$, and the effect sizes $\Delta_{\tau}$ and $\Delta_{\delta}$. In particular, we can see that the required number of networks, $K_{\tau}$ and $K_{\delta}$, increase with $\rho_{Y}$ and decrease with $n$.
Smaller effect sizes $\Delta_{\tau}$ and $\Delta_{\delta}$ also inflate the required sample sizes of index participants for testing the AIE and the ASpE with sufficient power, respectively.
From \eqref{K-td} and \eqref{K-o} we observe that $K_{J}$ and $K_{o}$ depend on $n$, $\rho_Y$, $p$ and the effect sizes of both $\tau$ and $\delta$. Smaller $\Delta_{\tau}$ and $\Delta_{\delta}$ inflate both $K_{J}$ and $K_{o}$. However, values $\Delta_{\tau}$ and $\Delta_{\delta}$ in different directions could result in a decrease or increase of $K_{J}$ and $K_{o}$. 
In practice with public health interventions, spillover effects are usually in the same direction as individual treatment effects.
Moreover, as $K_{\tau}$ and $K_{\delta}$, the required number of networks $K_{J}$ and $K_{o}$ grow with $\rho_{Y}$ and decrease with network size $n$. The proofs of these claims are provided in the Appendix \ref{DRV-K}.
We can also calculate the optimal $p$ to minimize the number of the networks, which satisfies the required power, depending on $n$ and $\rho_Y$ for testing AIE and ASpE, the formulas for calculating the optimal $p$ are given in Appendices \ref{DRV-K-T} and \ref{DRV-K-D}.% S.7.1 and S.7.2 of the Supplementary Materials. %Appendices \ref{DRV-K-T} and \ref{DRV-K-D}.
In %S.7.3 and S.7.4 of the Supplementary Materials, 
Appendices \ref{DRV-K-O} and \ref{DRV-K-TD}, 
we demonstrate that $p=0.5$ is the optimal value for testing
the overall effect, and also testing the AIE and the ASpE effect simultaneously for any $\rho_{Y}$ and $n$. We will evaluate the association between $K_C$ and the design parameters using numerical simulations because there is no closed form for $K_C$. 

\begin{figure}
\centerline{%
\includegraphics[width=\linewidth]{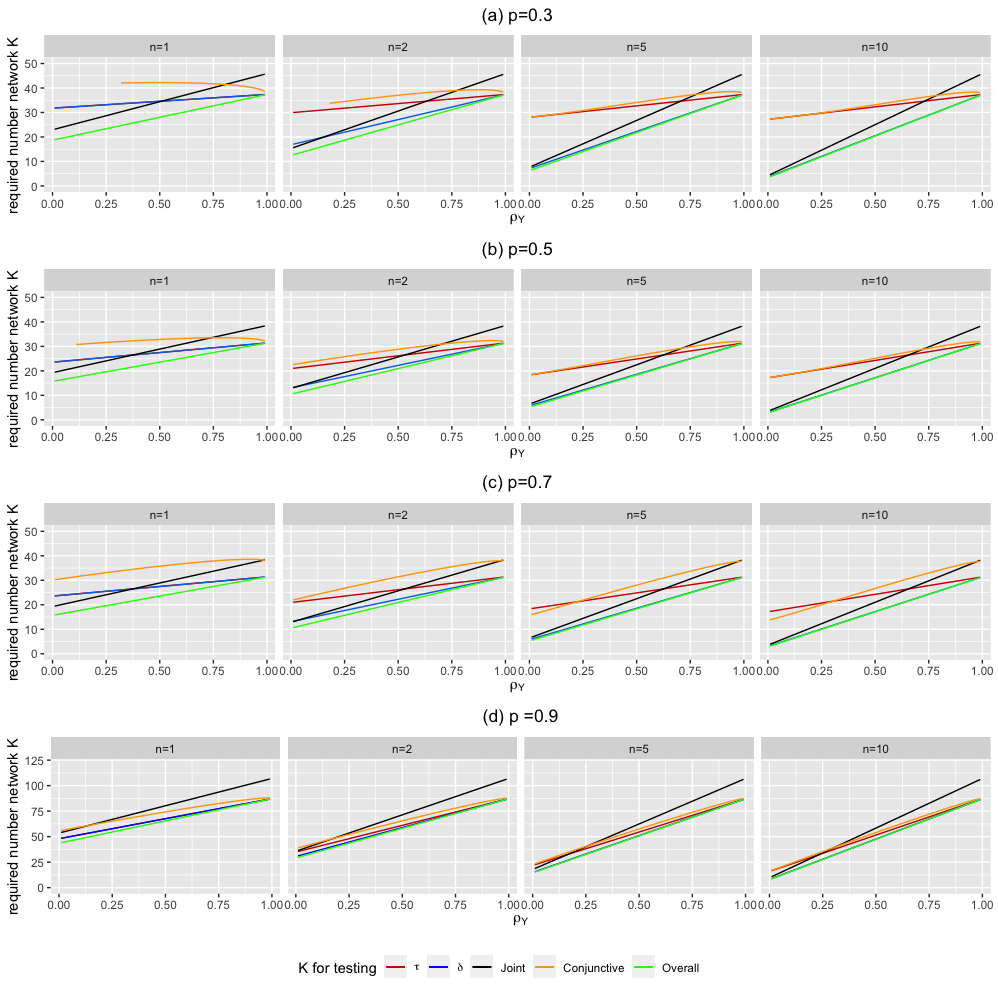}}
%% to include a figure, or to leave a blank space
\caption{Comparison of the required number of networks for HIE, HSpE, HISpJ, HISpC, and HOE given different values of $p$, $n$, and $\Delta_{\tau}=\Delta_{\delta}$=1, $\sigma^2_{Y} =1$. Note that with $n=1$ the blue and red lines overlap meaning that $K_{\tau}=K_{\delta}$.}
\label{fig:no_X_n}
\end{figure}

\begin{figure}
\centerline{%
\includegraphics[width=\linewidth]{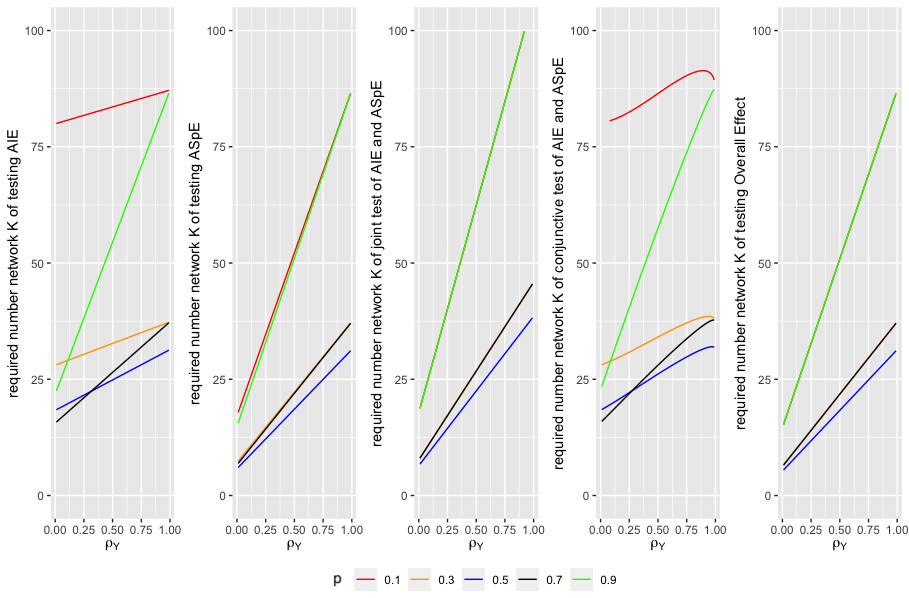}}
%% to include a figure, or to leave a blank space
\caption{Required number of index participants for HIE, HSpE, HISpJ, HISpC, and HOE given different values of $p$: $\Delta_{\tau}=\Delta_{\delta}$=1, $\sigma^2_{Y} =1$, $n=5$. The lines with two colors mean the two lines overlap with each other.}
\label{fig:no_X_p}
\end{figure}

We further investigate the role of $\rho_Y$, the number of network members $n$ and the effect sizes $\Delta_{\tau}$ and $\Delta_{\delta}$ in testing the AIE, the ASpE, the overall effect, and multiple effects for
the joint test and the conjunctive test using Model \eqref{model-1} with numerical simulations. 
For fixed $\sigma^2_Y$, $\Delta_{\tau}$ and $\Delta_{\delta}$ set to 1, we let the outcome ICC, $\rho_{Y}$, vary  from (0 to 1), $n \in \{1,2,5,10\}$, and $p = c(0.3,0.5,0.7,0.9)$. Figure \ref{fig:no_X_n} shows the required number of networks for the four hypothesis tests, HIE, HSpE, HISpJ, HISpC, and HOE  with $\alpha = 5\%$ and $\pi=80\%$. To note, there is no solution for HISpC in some parameter combinations (e.g., $p=0.3, n=1$ case), thus the lines in Figure~\ref{fig:no_X_n} are incomplete. 
For each $p$, the patterns of $K_{\tau}$, $K_{\delta}$, $K_{J}$, $K_{C}$, and $K_{o}$ with $\rho_Y$ and $n$ are displayed. Figure~\ref{fig:no_X_p} highlights how the required number of index participants for the different tests varies with $p$ and $\rho_Y$ given $n=5$.

\subsection{Comparison of power and sample size for the  hypotheses tests}

To further compare the required sample sizes to detect different effects, we fixed $p=0.5$, as it is a common choice, and compute the ratios between the required number of index participants for the different hypotheses. The ratio between $K_{\delta}$ and $K_{\tau}$ is
\begin{equation}
    K_{\delta/\tau} = \frac{K_{{\delta}}}{K_{{\tau}}} = \frac{(1-p)(1-\rho_Y)+n(1+n\rho_Y)}{n^2(1-p)(1-\rho_Y)+n(1+n\rho_Y)}\times  r_{\tau/\delta}^2,
    \label{K-d/t}
\end{equation}
where $r_{\tau/\delta} = \Delta_{\tau} / \Delta_{\delta} $ is the ratio between the AIE and the ASpE. 
As the individual effect becomes larger relative to the spillover effect, the number of index participants needed to assess the spillover effect becomes larger than the number needed to assess the individual effect.
%a larger $r_{\tau/\delta}$ results in a larger $K_{\delta/\tau}$.
When $r_{\tau/\delta} =1$, it is straightforward to see that $K_{\delta/\tau} \leq  1$, which means that detecting the spillover effect equal in magnitude to the individual effect requires fewer egonetworks than needed to detect the AIE. Intuitively, this is because the estimation of spillover effects relies on the comparison between $nKp$ network members versus $(n+1)K(1-p)$ untreated network members and index participants, whereas the estimation of the AIE relies on the comparison between $Kp$ treated index participants versus $(n+1)K(1-p)$ untreated individuals. That is, the ASpE has $n$ more participants available for estimation than the AIE. Thus, when $r_{\tau/\delta}=1$ and $n=1$, then $K_{\tau} = K_{\delta}$. Let $\rho_Y = 0$, we have
$
K_{\delta/\tau} = \frac{(1-p)+n}{n^2(1-p)+n}$, 
 This means that the ratio decreases as $n$ increases, as can be seen in \eqref{K-d/t}. When $n$ is large, we need many more index participants to obtain an adequately powered study for HIE than for HSpE. To investigate the relationship between $K_{\delta/\tau}$ and $\rho_Y$, we rewrite \eqref{K-d/t} as
$
K_{\delta/\tau} = \frac{\{n^2-(1-p)\}\rho_Y+n+1-p}{n^2p\rho_Y+n^2(1-p)+n} \times r_{\tau/\delta}
$.
For any fixed $n>1$ and $r_{\tau/\delta}$, $K_{\delta/\tau}$ decreases as $\rho_Y$ increases, which means $K_{\delta}$ and $K_{\tau}$ will get closer.  In practice, $\rho_Y$ is usually not that large, for example, $\rho_Y \approx 0.115$ in HPTN 037.

To compare $K_{J}$ with $K_{\tau}$ and $K_{\delta}$, we investigate the ratios:
\begin{equation}
 K_{J/\tau} = \frac{K_{J}}{K_{\tau}} = \frac{\upsilon(\chi^2_{1-\alpha}(2),\pi,2)}{(z_{1-\alpha/2}+z_{\pi})^2} \times \frac{(1+n\rho_{Y})(n+1)}{n(1-p)(1-\rho_Y)+(1+n\rho_Y)} \times r_{\tau/J}  
   \label{K-td/t}
\end{equation}
and
\begin{equation}
K_{J/\delta} = \frac{K_{J}}{K_{\delta}} = \frac{\upsilon(\chi^2_{1-\alpha}(2),\pi,2)}{(z_{1-\alpha/2}+z_{\pi})^2} \times \frac{(1+n\rho_{Y})(n+1)n}{(1-p)(1-\rho_Y)+n(1+n\rho_Y)} \times r_{\delta/J}, 
 \label{K-td/d}
\end{equation}
where $r_{\tau/J} = \frac{\Delta^2_{\tau}}{\Delta^2_{\tau}+n\Delta^2_{\delta}}$ and $ r_{\delta/J} = \frac{\Delta^2_{\delta}}{\Delta^2_{\tau}+n\Delta^2_{\delta}}$. It is clear that increasing $r_{\tau/J}$ inflates $K_{J /\tau}$. Since a larger $r_{\tau/J}$ indicates a larger $r_{\tau/\delta}$, we have that $K_{J/\tau}$ increases as the effect size ratio between $\tau$ and $\delta$ increases. We also observe that $ r_{\delta/J}$ inflates $K_{J/\tau}$. Since $r_{\delta/J}$ increases as $r_{\tau/\delta}$ decreases, this confirms that $K_{J/\delta}$ increases as the effect size ratio between $\tau$ and $\delta$ decreases, that is, as the effect of spillover becomes larger relative to the individual treatment effect. When $\Delta_{\tau} =\Delta_{\delta}$, then \eqref{K-td/t} simplifies to
$
K_{J /\tau} = \frac{\upsilon(\chi^2_{1-\alpha}(2),\pi,2)}{(z_{1-\alpha/2}+z_{\pi})^2} \times \frac{(1+n\rho_{Y})}{n(1-p)(1-\rho_Y)+(1+n\rho_Y)}
$
with $\upsilon(\chi^2_{1-\alpha}(2),\pi,2) /(z_{1-\alpha/2}+z_{\pi})^2 \approx 1.227$ for $\alpha = 0.05$.
When $\rho_Y = 0$, we then have
$
K_{J/\tau} =  1.227/\{n(1-p)+1\}.
$
We observe that for any $n$, $K_{J} < K_{\tau}$, which means that detecting the AIE requires more index participants than detecting the AIE and the ASpE simultaneously, and the ratio decreases as $n$ increases. This indicates that HIE is more sensitive to $n$ than HISpJ. 
With fixed $n$, 
%the ratio can be wrote as
%\[
%r_{\tau\delta /\tau} = 1.227 \times \frac{n\rho_Y+1}{np\rho_Y+n-np+1},
%\]then 
as $\rho_Y$ increases, $r_{J /\tau}$ increases. This indicates that $K_{J}$ is more sensitive to $\rho_{Y}$ then $K_{\tau}$.

Similar to $K_{J /\tau}$, when $\Delta_{\tau} =\Delta_{\delta} $, we write \eqref{K-td/d} as
$
K_{J/\delta} = 1.227 \times \frac{n(1+n\rho_{Y})}{(1-p)(1-\rho_Y)+n(1+n\rho_Y)}.
$
When $n>2$, $K_{J/\delta}$ is larger than 1. This means that we need more egonetworks to detect the AIE and the ASpE simultaneously than to detect the ASpE only when the index participant has more than two network members. When $n=1$ or 2, as $\rho_{Y}$ becomes larger, $K_{J}$ is first smaller than $K_{\delta}$, and then larger than $K_{\delta}$. %we have $K_{\tau \delta}<K_{\delta}$.
Furthermore, given $n$, it is clear that an increase in $\rho_Y$ inflates $K_{J/\delta}$, as it dose with $K_{\tau}$, $K_{\delta}$ and $K_{J}$. This indicates that the increasing rate of $K_{J}$ as a function of $\rho_{Y}$ is greater than that of $K_{\delta}$.  When $\rho_Y = 0$, 
$
K_{J/\delta} =  1.227 \times \frac{n}{n-p+1},
$
which means $K_{J/\delta} $ increases as $n$ increases. It means that HISpJ depends more on $n$ than HSpE.

Last but not least, we compare $K_{\tau}$, $K_{\delta}$, and $K_{J}$ with $K_o$. We investigate the following ratios:
\begin{equation}
\begin{aligned}
    K_{\tau/o} = \frac{K_{\tau}}{K_o} = \frac{n(1-p)(1-\rho_{Y})+(1+n\rho_{Y}) }{(n+1)^2 (1+n\rho_Y)} \times \left(1+\frac{2n}{r_{\tau/\delta}}+\frac{n^2}{r^2_{\tau/\delta}}\right),
\end{aligned}
\label{K-t/o}
\end{equation}
 \begin{equation}
\begin{aligned}
    K_{\delta/o} = \frac{K_{\delta}}{K_o} = \frac{(1-p)(1-\rho_{Y})+n(1+n\rho_{Y}) }{n(n+1)(1+n\rho_Y)} \times \left( r^2_{\tau/\delta} +2n r_{\tau/\delta}  +n^2\right)
\end{aligned}
\end{equation}
 and 
 \begin{equation}
\begin{aligned}
    K_{J/o}=\frac{K_{J}}{K_o}  =\frac{\upsilon(\chi^2_{1-\alpha}(2),\pi,2)}{(n+1)(z_{1-\alpha/2}+z_{\pi})^2}  \times \frac{ r^2_{\tau/\delta}+2n r_{\tau/\delta}+n^2}{ r^2_{\tau/\delta}+n}  .
\end{aligned}
\label{K-td/o}
\end{equation}
From \eqref{K-t/o}-\eqref{K-td/o}, we see that $K_{\tau/o}$ could either be larger or smaller than 1 depending on the magnitude of other parameters. We also observe that $K_{\tau/o}$ decreases as $p$, $\rho_Y$, or $r_{\tau/\delta}$ increasing. But $K_{\delta/o}$ is always larger than 1, which means that we need more egonetworks to test ASpE than OE given other parameters fixed. Similar to  $K_{\tau/o}$, $K_{\delta/o}$ decreases as $p$ and $\rho_Y$ increases, but increasing $r_{\tau/\delta}$ inflates $K_{\delta/o}$. For $K_{J/o}$, when $n>r_{\tau/\delta}$, $K_{J/o}$ increases as $n$ increases; when $n<r$, $K_{J/o}$ decreases as $n$ increases. When $r_{\tau/\delta}<1$, $K_{J/o}$ increases as $r_{\tau/\delta}$ increases; when $r_{\tau/\delta}>1$, $K_{J/o}$ decreases as $r_{\tau/\delta}$ increases.

\section{Illustrative Example: HPTN 037 Study}
\label{data}

We use 
%data collected from 
the HPTN 037 study to inform the design parameters and show how to estimate the required number of index participants given pre-specified power and level of significance for different hypothesis tests.
The HPTN 037 study is an ENR experiment to assess the efficacy of a network-oriented peer education intervention to promote HIV risk reduction among injection drug users and their drug network members in Chiang Mai, Thailand, and Philadelphia, US. 
Here, 
%following the recommendation of the study investigators \citep{Latkin2009}, 
we only include participants in the Philadelphia site \citep{Latkin2009}. 
%due to the profound effect of a governmental policy known as the “war on drugs” happened in Thailand.
%Participants were followed up 30 months after the baseline visit with biannual visits to obtain information on risk behaviors of drug injections. 
At the baseline visit, eligible index participants were those who injected drugs at least 12 times in the prior three months, and eligible network members were those who injected drugs with the relevant index participant within the prior three months.
%All participants received Voluntary HIV counseling and testing at each study visit. 
Index participants received network-oriented peer educator training sessions during a four-week period and two booster sessions at six and 12 months of study participation.

%We use HIV Prevention Trials Network 037 (HPTN 037) \citep{Latkin2009} to inform the design parameters and demonstrate how to estimate the required number of index participants given pre-specified power and Type I error for the five hypothesis tests considered in this paper. \textcolor{blue}{}
%In this section, following the recommendation of the study investigators \citep{Latkin2009}, we include participants from the Philadelphia site alone due to the co-occurrent governmental policy known as the “war on drugs” which occurred in Thailand at the same time the study was concluded, rendering the data at that site less reliable. 

The primary outcome considered here is the average number of drug injection risk behaviors in the month prior to each visit, averaged across the number of visits each individual attended up to the 30-month visit. Here, we consider ``using rinse water that others had used'' as the risk injection behavior to illustrate our approach. 
%We observed that some index participants and network members had some extreme values of this risk injection behavior up to 150 times. 
To make the estimation result more robust, we removed the index participants and the network members whose average number of risk injection behaviors lie more than 1.5 times the interquartile range.
%After removing outliers in the primary outcomes, 186 networks remained. 
The remaining 186 networks 
%have at least 1 network member and at most 6 network members.
have a number of network members ranging between 1 and 6, with an average of 2 (sd=1.15).
%The average number of network members was 2
%with a standard error of 1.15. 
The intervention assignment probability, $p$, was 0.47, with 88 intervention egonetworks and 98 control egonetworks. The outcome variance and the outcome ICC were $\sigma_Y^2=1.02$ and $\rho_Y  \approx 0.115$, with an outcome mean of 0.53. The ASpE and the the AIE were estimated to be $-0.34$ (95\% CI:$[-0.61, -0.08]$) and $-0.32$ (95\% CI:$[-0.63, -0.01]$), respectively. This means that the average number of drug injection risk behaviors of the treated index participants is 0.32 less than that of the untreated index participants, 
%after the peer education interventions, 
and the average number of drug injection risk behaviors of the network members whose index participants were treated is 0.34 less than that of the network members whose index participants were not treated. We approximately follow these features of HPTN 037 to set the design parameters.
%, we set AIE and ASpE to -0.35, the number of index participants to 186, and the number of network members to 2, $\rho_Y$ to 0.1, and $p$ to 0.5.
Table~\ref{t:tab1} shows the required number of egonetworks needed for HIE, HSpE, HISpJ, HISpC, and HOE, to ensure 80\% power to detect the effects, with number of network members equal to 2,  with different effect sizes, $p$, and $\rho_Y$.
%The effect size of the AIE is very close to zero in HPTN 037 study but not siginificant Usually the AIE is larger than the ASpE. 
We varied these design parameters approximately around the values estimated in HPTN 037 to assess the sensitivity of $K$ to these changes.
In particular, we set the the effect size of both the AIE and the ASpE to 0.5 times, 1 times, and 1.5 times the value of -0.35, approximately the value estimated in HPTN 037.
%We also set the effect size of the AIE to be the same, and 1.5 times the value of AIE estimated in HPTN 037.   
We also varied $p$ and $\rho_Y$ using the values
$[0.5, 0.3, 0.7]$ and $[0.1, 0.2, 0.05]$, respectively.
Using equations \eqref{K-tau}-\eqref{K-o} to calculate $K_{\tau}$, $K_{\delta}$, $K_{J}$, $K_{C}$ and $K_{o}$, we found that when $\rho_{Y}=0.10$ and $p=0.50$, $K_{\delta} =122$ and $K_{\tau} = 251$ egonetworks are required to ensure $80\%$ power to detect the ASpE and AIE with sizes similar to that observed in HPTN 037, i.e., $\Delta_{\delta}=-0.35$ and $ \Delta_{\tau}=-0.35$. To detect either the AIE or the ASpE, the AIE and the ASpE simultaneously and the overall effect with $80\%$ power, we need $K_{J}=126$, $K_{C}=195$ and $K_o =103$, respectively. When we shifted the design parameters, we observed similar patterns as shown in Figure \ref{fig:no_X_n}. In summary, given the parameters estimated from HPTN 037, the HOE requires the smallest number of networks, and the HISpJ requires the largest number of networks. Increasing assignment probability $p$ or decreasing outcome ICC requires less index participants for all the hypothesis tests. Amplifying the effect sizes reduces the required number of index participants to detect the causal effects.

\begin{table}
  \centering
   \def\~{\hphantom{0}}
   \begin{minipage}{19cm}
   \caption{Required number of networks for Hypothesis Test 1-4 with $n=2$, and $\sigma^2_Y=1.02$  \\ in HPTN 037}  \label{t:tab1}%
   \resizebox{0.90\textwidth}{!}{%
    \begin{tabular*}{\textwidth}{ccccccccccccccccccc}
     \toprule
    \multicolumn{2}{c}{\multirow{2}[3]{*}{ Parameter}} & \multicolumn{5}{c}{$\Delta_{\delta}=-0.350$} &       & \multicolumn{5}{c}{$\Delta_{\delta}=-0.175$} &       & \multicolumn{5}{c}{$\Delta_{\delta}=-0.525$} \\
\cline{3-7}\cline{9-13}\cline{15-19}  
\multicolumn{2}{c}{} & $K_{\delta}$ & $K_{\tau}$ & $K_{J}$ & $K_{C}$ & $K_{o}$ &     & $K_{\delta}$ & $K_{\tau}$ & $K_{J}$ & $K_{C}$ & $K_{o}$ &      & $K_{\delta}$ & $K_{\tau}$ & $K_{J}$ & $K_{C}$ & $K_{o}$ \\
     \hline
    $\rho_{Y}$ & $p$   & \multicolumn{5}{c}{Case 1: $\Delta_{\tau}=-0.350$} &         & \multicolumn{5}{c}{Case 1: $\Delta_{\tau}=-0.350$} &       &  \multicolumn{5}{c}{Case 1: $\Delta_{\tau}=-0.350$} \\
\hline   \multirow{3}[6]{*}{0.1} & 0.5   & 122   & 180   & 126   & 195   & 103   &        & 487   & 180   & 252   & 489   & 231   &         & 55    & 180   & 69    & 180   & 58 \\
\cline{2-7}\cline{9-13}\cline{15-19}         & 0.3   & 155   & 251   & 153   & 268   & 123   &       & 617   & 251   & 300   & 622   & 275   &        & 69    & 251   & 82    & 251   & 69 \\
\cline{2-7}\cline{9-13}\cline{15-19}         & 0.7   & 136   & 177   & 150   & 195   & 123   &     & 544   & 177   & 300   & 544   & 275   &       & 61    & 177   & 82    & 178   & 69 \\
\cline{1-7}\cline{9-13}\cline{15-19}    \multirow{3}[6]{*}{0.2} & 0.5   & 137   & 188   & 147   & 206   & 120   &      & 547   & 188   & 294   & 548   & 270   &     & 61    & 188   & 81    & 189   & 68 \\
\cline{2-7}\cline{9-13}\cline{15-19}          & 0.3   & 171   & 257   & 175   & 278   & 143   &      & 684   & 257   & 350   & 686   & 321   &       & 76    & 257   & 96    & 257   & 81 \\
\cline{2-7}\cline{9-13}\cline{15-19}         & 0.7   & 155   & 192   & 175   & 211   & 143   &      & 619   & 192   & 350   & 619   & 321   &     & 69    & 192   & 96    & 192   & 81 \\
\cline{1-7}\cline{9-13}\cline{15-19}     \multirow{3}[6]{*}{0.05} & 0.5   & 115   & 176   & 116   & 189   & 94    &     & 458   & 176   & 231   & 460   & 212   &    & 51    & 176   & 63    & 176   & 53 \\
\cline{2-7}\cline{9-13}\cline{15-19}        & 0.3   & 146   & 248   & 138   & 263   & 112   &        & 583   & 248   & 275   & 591   & 252   &       & 65    & 248   & 75    & 248   & 63 \\
\cline{2-7}\cline{9-13}\cline{15-19}       & 0.7   & 127   & 170   & 138   & 186   & 112   &       & 506   & 170   & 275   & 506   & 252   &       & 57    & 170   & 75    & 170   & 63 \\
  \hline
          &       & \multicolumn{5}{c}{Case 2: $\Delta_{\tau}=-0.525$} &       & \multicolumn{5}{c}{Case 2: $\Delta_{\tau}=-0.525$} &  &\multicolumn{5}{c}{Case 2: $\Delta_{\tau}=-0.525$} \\
    \hline
    \multirow{3}[6]{*}{0.1} & 0.5   & 122   & 80    & 89    & 131   & 76    &       & 487   & 80    & 138   & 487   & 148   &     & 55    & 80    & 56    & 87    & 46 \\
\cline{2-7}\cline{9-13}\cline{15-19}         & 0.3   & 155   & 112   & 106   & 173   & 90    &       & 617   & 112   & 164   & 617   & 176   &      & 69    & 112   & 67    & 120   & 55 \\
\cline{2-7}\cline{9-13}\cline{15-19}         & 0.7   & 136   & 79    & 106   & 140   & 90    &      & 544   & 79    & 164   & 543   & 176   &       & 61    & 79    & 67    & 87    & 55 \\
\cline{1-7}\cline{9-13}\cline{15-19}    \multirow{3}[6]{*}{0.2} & 0.5   & 137   & 84    & 104   & 144   & 88    &     & 547   & 84    & 161   & 547   & 173   &       & 61    & 84    & 66    & 92    & 54 \\
\cline{2-7}\cline{9-13}\cline{15-19}          & 0.3   & 171   & 114   & 124   & 185   & 105   &        & 684   & 114   & 191   & 684   & 206   &      & 76    & 114   & 78    & 124   & 64 \\
\cline{2-7}\cline{9-13}\cline{15-19}          & 0.7   & 155   & 85    & 124   & 158   & 105   &       & 619   & 85    & 191   & 618   & 206   &     & 69    & 85    & 78    & 94    & 64 \\
\cline{1-7}\cline{9-13}\cline{15-19}     \multirow{3}[6]{*}{0.05} & 0.5   & 115   & 78    & 82    & 125   & 70    &    & 458   & 78    & 126   & 458   & 136   &       & 51    & 78    & 52    & 84    & 42 \\
\cline{2-7}\cline{9-13}\cline{15-19}          & 0.3   & 146   & 110   & 97    & 167   & 83    &      & 583   & 110   & 150   & 583   & 162   &      & 65    & 110   & 62    & 117   & 50 \\
\cline{2-7}\cline{9-13}\cline{15-19}       & 0.7   & 127   & 76    & 97    & 132   & 83    &       & 506   & 76    & 150   & 506   & 162   &       & 57    & 76    & 62    & 83    & 50 \\
\hline          &       & \multicolumn{5}{c}{Case 3: $\Delta_{\tau}=-0.70$} &         & \multicolumn{5}{c}{Case 3: $\Delta_{\tau}=-0.70$} &      & \multicolumn{5}{c}{Case 3: $\Delta_{\tau}=-0.70$} \\
\hline    \multirow{3}[6]{*}{0.1} & 0.5   & 122   & 45    & 63    & 123   & 58    &     & 487   & 45    & 84    & 487   & 103   &      & 55    & 45    & 45    & 63    & 37 \\
\cline{2-7}\cline{9-13}\cline{15-19}           & 0.3   & 155   & 63    & 75    & 156   & 69    &       & 617   & 63    & 100   & 617   & 123   &     & 69    & 63    & 53    & 85    & 44 \\
\cline{2-7}\cline{9-13}\cline{15-19}         & 0.7   & 136   & 45    & 75    & 136   & 69    &       & 544   & 45    & 100   & 543   & 123   &      & 61    & 45    & 53    & 66    & 44 \\
\cline{1-7}\cline{9-13}\cline{15-19}     \multirow{3}[6]{*}{0.2} & 0.5   & 137   & 47    & 74    & 137   & 68    &      & 547   & 47    & 98    & 547   & 120   &       & 61    & 47    & 52    & 68    & 44 \\
\cline{2-7}\cline{9-13}\cline{15-19}          & 0.3   & 171   & 65    & 88    & 172   & 81    &     & 684   & 65    & 117   & 683   & 143   &        & 76    & 65    & 62    & 89    & 52 \\
\cline{2-7}\cline{9-13}\cline{15-19}         & 0.7   & 155   & 48    & 88    & 155   & 81    &       & 619   & 48    & 117   & 618   & 143   &      & 69    & 48    & 62    & 73    & 52 \\
\cline{1-7}\cline{9-13}\cline{15-19}    \multirow{3}[6]{*}{0.05} & 0.5   & 115   & 44    & 58    & 115   & 53    &    & 458   & 44    & 77    & 457   & 94    &        & 51    & 44    & 41    & 61    & 34 \\
\cline{2-7}\cline{9-13}\cline{15-19}         & 0.3   & 146   & 62    & 69    & 148   & 63    &        & 583   & 63    & 92    & 583   & 112   &        & 65    & 62    & 49    & 82    & 41 \\
\cline{2-7}\cline{9-13}\cline{15-19}        & 0.7   & 127   & 43    & 69    & 127   & 63    &        & 506   & 43    & 92    & 505   & 112   &        & 57    & 43    & 49    & 62    & 41 \\
     \bottomrule
    \end{tabular*}%
    }
    \end{minipage}
\end{table}%

\vspace{-0.3cm}

\section{Conclusion and Discussion}
\label{con}
The no-interference assumption is naturally violated by participants with social or physical interactions in a network. 
%The influence of one unit's intervention on other units through the network affects the treatment's effect on the participants, but 
Few studies have been able to estimate the ASpE with sufficient power. The lack of methods for designing studies to detect the ASpE motivates us to propose an egonetwork-based design in which only a single index participant in treated networks receives the intervention. In this ENR design, we can estimate the AIE, the ASpE, and the overall effect simultaneously with a regression model.
 We developed closed-form sample size formulas for calculating the required number of index participants to power the hypothesis tests of detecting several causal estimands of interest. We investigated the patterns of how the required number of index participants changes with design parameters.
 %accounting for shifting outcome intra-class correlation coefficients and treatment assignment probability for the AIE, ASpE, and overall effect. 
 %Our procedures provide a simple model-based approach to identify the AIE, ASpE, and the overall effect with the closed-form sample size formulas. 
 In addition, we considered an alternative model to estimate and test the AIE and the ASpE using two separate linear mixed effect models in the Appendix \ref{AlterModel}. %Supplementary Material S.9.
 Comparing to the model proposed in Section \ref{Regression}, the alternative model allows the outcomes of index participants and their network members to have different total variances, but it requires greater index participants for testing the AIE and ASpE.

This work is affected by a few limitations. The study design assumes the same number of members in each network. Our methods could be further extended to handle variable network sizes, as in \citet{Manatunga2001}. In HPTN 037, for example, the mean network size was 2, but it varies from 1 to 6, thus including the variation of the network size would possibly change the required number of networks.
Furthermore, we used asymptotic statistical results to construct the sample size formulas, and their accuracy in finite samples could be further studied. 
%Future work could develop the design of studies investigating interventions package while here we only considered a single intervention \citep{Buchanan2022}. Packages of several interventions may significantly increase the intervention effect \citep{Padian2011,DeGruttola2010}, but the best way to develop and evaluate the interventions package in the presence of spillover is unknown. Egocentric network-based design can be extended to solve this problem. 
We could also consider index and/or individual covariates that may lead to heterogeneity of the spillover effect. Design of studies to assess heterogeneity in the ASpE, AIE, and AOE will be considered in future work. To facilitate the use of our proposed design, we will also develop software for sample size calculation and power analysis.
%, such as R packages.

%\section{Supplementary Material}
%\label{supple}
%In the supplementary material, we provide the details of remarks, the proofs of the assumptions and sample size calculation formulas, the identification of causal estimands in the manuscript.

% \href{http://biostatistics.oxfordjournals.org}%
% {http://biostatistics.oxfordjournals.org}.
%\url{http://biostatistics.oxfordjournals.org}.

%\backmatter

%%%%%%%%%%%%%%%%%%%%%%%%%%%%%%%%
% %ADD AGAIN IN THE FINAL VERSION
% \section*{Supporting Information}

% \label{supple}
% In the supplementary material, we provide the details of remarks, the proofs of the assumptions and sample size calculation formulas, the identification of causal estimands in the manuscript.

 \section*{Acknowledgement}

 This research was partially supported by the Avenir Award Program for Research on Substance Abuse and HIV/AIDS (DP2) from National Institute on Drug Abuse of the National Institutes of Health (DP2DA046856).
%%%%%%%%%%%%%%%%%%%%%%%%%%%%%%%%%%%%%%%%%%%%%%%%%%%%%%%%%%%%%%%%%%%%%%%%%%%

\clearpage

%\bibliographystyle{biom}
%\bibliography{ref.bib}

%%%%%%%%%%%%%%%%%%%%%%%%%%%%%%%%%%%%%%%%%%%%%%%%%%%%%%%%%%%%%%%%%%%%%%%%%%%

\clearpage

\section{Appendix}
\subsection{Population Notation and Egocentric Notation}
\label{app:notation_remarks}

\subsubsection{Population Notation}
%\subsubsection{Remark 1: Transformation from Single-index notation to Double-index notation}

Let us denote by $\mathcal{M}$ a population of interest and by $\mathcal{G}$ an undirected network being the graphical representation of this population, which consists of the pair $(\mathcal{M}, E)$, where $\mathcal{M}$ is the set of units (or nodes) and $E$ is the 
%$E=\{e_{mq}\}_{m,q\in \mathcal{M}}$ is a 
set of edges representing the links between two units in $\mathcal{M}$. 
In Section \ref{sec:notation}, we have denoted individuals by $ik$, where $k$ is the egonetwork indicator. However, in a ENT design, individuals are sampled from a population $\mathcal{M}$ using an egocentric strategy, where we first sample index participants and then information is collected on their network members. Therefore, egonetworks are entities that are generated after sampling. In fact, in the population $\mathcal{M}$ each individual have their own egocentric networks (or network neighborhood), regardless of whether they are going to be sampled as an index participant, a network member, or not sampled at all. In addition, egocentric networks of individuals overlap, given that multiple individuals can share the same network neighbor (e.g., friend or sex partner). For this reason, the egonetwork notation $ik$ cannot be used in the population, for which we must use a single index notation $m$, such that $\mathcal{M}=\{m\}_1^M$. In a ENT design, we can also denote by $\mathcal{R}\subset\mathcal{M}$ the set of index participants. 

Using the population notation, the set of edges $E$ in the graph $\mathcal{G}$ can be represented by $E=\{e_{mq}\}_{m,q\in \mathcal{M}}$, where each edge $e_{mq}$ represents a link between two units $m$ and $q$. We now denote by 
%
%Let us denote by $\mathcal{M}$ a population of interest and by $\mathcal{G}$ an undirected network being the graphical representation of this population, which consists of the pair $(\mathcal{M}, E)$. $E=\{e_{mq}\}_{m,q\in \mathcal{M}}$ is a set of edges representing the links between two units in $\mathcal{M}$. 
%Given a graph $\mathcal{G}$, we denote by
$\mathcal{N}_m=\{q\in \mathcal{M}: e_{mq}\in E\}$ the set of units sharing a link with unit $m$, i.e., the `network neighbors' of unit $m\in \mathcal{R}$. 
%%
%
%The networks of the population are constructed by the connected individuals.
% In an egocentric network-based randomized study,
% a set of index participants $\mathcal{R}\subset\mathcal{M}$ is sampled from the population, $\mathcal{M}$, and randomly assigned to an intervention.
% %with probability $p$, with $0<p<1$. 
% In addition, these index participants are asked to provide a list of individuals that they consider to be members of their social network (e.g., friends, sexual partners, drug use partners), i.e.,  
Then,
the study sample $\mathcal{N}\subset\mathcal{M}$ consists of all the index participants and their social network members, i.e., $\mathcal{N}=(\mathcal{R},\bigcup_{m\in \mathcal{R}} \mathcal{N}_{m})$. 
%Network members are not directly given the intervention. Information about baseline characteristics and the outcome of interest is collected for both index participants and network members as part of the baseline and follow-up surveys.
%For consistency with the network literature \citep{egonetwork}, we call `egos' the index participants and `alters' the network members.  
%The set of alters of an ego is commonly referred to as `egocentric network', and here we call `(ego)network' the set of network members of an index participant together with the index participant itself.
%In addition, we call intervention (ego)networks the egonetworks with treated egos, and control (ego)networks the egonetworks where the ego is not treated. 

In Section \ref{sec:notation}, 
we define the variables of interest, including the treatment, using the egonetwork notation.
Using the population notation we can denote by $Z_m$ the treatment of unit $m\in \mathcal{M}$. The assumption that all units that are not in the sample are considered as not treated because they cannot receive the intervention implies that $Z_{m}=0, \forall m \notin \mathcal{N}$. 
%Note that all units that are not in the sample are considered as not treated, that is, $Z_{m}=0\, \forall m\notin \mathcal{N}$. 
We can also denote by 
$G_{m}$ the number of treated network neighbors of unit $m$, i.e.,  $G_{m}=\sum_{q \in \mathcal{N}_{m}} Z_{m}$.
However, because we assume that individuals that are not in the sample cannot receive the treatment,
$G_{m}=\sum_{q \in \mathcal{N}_{m}\cap \,\mathcal{N}} Z_{m}$.

\subsubsection{Relation between the Population Notation and Egocentric Notation}

% In Section 2.1\ref{sec:notation}, 
% we denote by $\mathcal{M}$ the population of interest and by $\mathcal{N}_m$ the set of network `neighbors' or social network members of a unit $m\in \mathcal{M}$. In an egocentric network-based design, 
% we define the set of index participants as $\mathcal{R}$ and the study sample $\mathcal{N}\subset\mathcal{M}$ consists of all the index participants and their social network members, i.e., $\mathcal{N}=(\mathcal{R},\bigcup_{m\in \mathcal{R}} \mathcal{N}_{m})$.  
In Section \ref{sec:notation}, for the sake of simplicity, we relabel the units in the sample $\mathcal{N}$ with the notation commonly used with clustered data. 
We let $k=1,...,K$ be the egonetwork indicator in $\mathcal{N}$. We then let $ik$ be the unit $i$ in egonetwork $k$, with $i=1, \ldots, n_k+1$. 
In particular, we let $1k$ represent the $k$th index participant and $ik$, with $i=2,..., n_k+1$, represents a network member of the egonetwork $k$. We refer to this notation as the egonetwork notation.
%With a slight abuse of notation, we denote by $\mathcal{N}_{ik}$ the network neighborhood of unit $i$ in egonetwork $k$.
%In this subsection, we will explain the relations between these two notation systems.

The relation between the population notation and the  egonetwork notation can be explained by mapping functions.
The egonetwork indicator $k$ is such that 
$m=r(k)\in \mathcal{R}$, with $r(\cdot)$ being some function mapping the $k$-th index participant to a unit in the population $\mathcal{R}$.
Hence,
$\mathcal{N}_{r(k)}$ represents the set of network members of the $k$-th index participant of size $n_k=|\mathcal{N}_{r(k)}|$. Similarly, $q=n(ik)\in \mathcal{N}_{r(k)}$ is the population indicator for $i$-th network member in the egonetwork $k$, with $n(\cdot)$ being some function mapping the $i$-th network member of the $k$-th index participant to a unit in the population $\mathcal{M}$. In general, we have that $m=h(ik) \in \mathcal{N}$ is the population indicator corresponding to the sample unit $ik$, where $h(1k)=r(k)$ and $h(ik)=n(ik)$ for $i=1, \dots, n_k$.
%More formally, using the population indexing, $\mathcal{N}_{ik}$ should be written as $\mathcal{N}_{h(ik)}$.

\subsubsection{Non-overlapping Egonetworks Assumptions}

In Section \ref{sec:nonoverlap}, 
we have introduced the non-overlapping egonetworks assumption (Assumption  \ref{ass:nonoverlap}) under this egonetwork notation.
% , we denote by $\mathcal{N}^*=\{ik\}_{k=1, \dots, K; \,i=1, \ldots, n_k}$ our sample of units, and by $\mathcal{N}^*_k=\{ik \in \mathcal{N}\}_{i=1, \dots, n_k}$ the subsample within each egonetwork $k$.
% Finally, let $R_{ik}$ be and indicator for whether unit $ik$ is an index participant ($R_{ik}=1$) or a network member ($R_{ik}=0)$. Given our egocentric notation, it follows that $R_{1k}=1$ and $R_{ik}=0$ for all $i>1$.
%  We denote by $\mathcal{N}^*_{ik}\subset \mathcal{N}^*$ the network neighborhood of unit $i$ in egonetwork $k$ only including in-sample units.
% Then, we make Assumption \ref{ass:nonoverlap}: $\mathcal{N}^*_{ik}\subset \mathcal{N}^*_{k} \quad \forall ik \in \mathcal{N}^*$.
Using the population indexing, we can formalize the non-overlapping egonetworks assumption in an alternative way. We denote by 
%$\mathcal{C}_k=\{h(ik), i=1, \dots, n_k\}$ 
$\mathcal{C}_k=r(k)\cup \mathcal{N}_{r(k)}$
the set of units in the egonetwork $k$, including both the index participant and the network members. The non-overlapping assumption can be formalized by assuming that
$\bigcap_{k=1}^K \mathcal{C}_k= \emptyset$. This assumption implies that 
$\mathcal{N}_{h(ik)}\bigcap \mathcal{C}_{k'} = \emptyset$, that is, the network neighborhood of a unit $h(ik)\in \mathcal{N}$ may include units that are not in the sample but cannot include those that are in other egonetworks.
The non-overlapping egonetworks assumption can be guaranteed by design by selecting index participants whose egocentric networks are unlikely to overlap. For example, we could use a block (or stratified) design. In particular, we would divide the sample into clusters (e.g., disparate geographical areas) such that units of different clusters are not connected, and sample only one index participant per cluster.

% using $Y_{ik}$, the outcome variable for individual $i$ in egonetwork $k$, $Z_{ik}$, the treatment variable. We make an assumption that
% in an egocentric network-based randomization design, only index participants can be treatment while network members cannot, i.e.,
% $Z_{1k}\in \{0,1\}$ and $Z_{ik}=0$ for $i>1$. 
% We also assume that all units that are not in the sample are considered as not treated because they cannot receive the intervention.
% Using the population indexing, this assumption can be written as follows: $Z_{m}=0 , \forall m \notin \mathcal{N}$.

% \begin{remark}[3: Overall Effect]

% %\subsubsection{Remark 4: Overall Effect}
% In Section 4.2, we show the causal estimatand expression of AIE, ASpE and AOEoverall effect. However
% in our egocentric network-based design, the total effect, which would be defined by the comparison $\E[Y(1,1)-(0,0)]$ \citep{HudgensHalloran2008}, is not identifiable here since there is no unit that is treated and has one network neighbor treated at the same time.

% \end{remark}

\subsubsection{Neighborhood Interference and Stratified Interference}
It is worth mentioning the relationship between the neighborhood interference assumption (Assumption \ref{ass:neighint})
and the common stratified interference assumption \cite{HudgensHalloran2008}. The latter rules interference between groups allowing interference only between groups (partial interference) and assumes that interference depends only on the number of treated individuals in the same group and not on who they are \citep{HudgensHalloran2008,Buchanan2018}.
Under the specific egocentric network-based design, where only index participants can be treated, and  
under the non-overlapping egonetworks assumption, the neighborhood interference assumption (Assumption \ref{ass:neighint}) 
 is equivalent to the stratified interference assumption with groups being the egonetworks. In fact, as already shown, $G_{ik}$, which is by definition equal to the number of network neighbors, i.e., $G_{ik}=\sum_{\ell k \in \mathcal{N}^*_{ik}} Z_{\ell k} $, in this setting will also be equal to the number of the other individuals in the same egonetwork who are treated, i.e., $G_{ik}=\sum_{\ell k \in \mathcal{N}^*_k, \ell\neq i} Z_{\ell k} $.
In the egocentric network-based study, $Z_{ik}$ and $G_{ik}$ can only take value 0 or 1 since at most one participant can be treated in an egonetwork.

% \subsection*{S.1.5 Random Sampling}

% In Section 3.3, %\ref{sec:identification}, 
% we introduce Assumption 4:
% %\ref{ass:ran_sam}: 
% $Y_{ik}(z, g) \perp R_{ik}$, which
% states that potential outcomes do not depend on whether the unit is an index participant or a network member, that is, index participants are randomly sampled from the population and network members can also be seen as such.
% Assumption 4 
% %\ref{ass:ran_sam} 
% could be problematic if index participants self-select themselves to serve in this role. In this case, index participants and network members may differ in terms of their characteristics. If this self-selection is a concern,  one can estimate the individual effect only among index participants and the spillover effect only among network members. In this case, the comparison group for the estimation of the individual effect would only include untreated index participants, while the comparison group for the estimation of the spillover effect would only include network members of untreated index participants.

\subsection{Identifying Assumptions}
%\subsection{Identifying Assumption and Identification of Causal Effects}
\label{sec:identification_assumptions}

%% MOVED TO THE APPENDIX
Under the randomization scheme of the ENR design, the following unconfoundedness assumption holds:

\begin{assumption}[Unconfoundedness of the treatment in the egocentric network-based design]
\label{ass:unc_ego}
$Y_{ik}(z, g) \perp Z_{ik} | R_{ik}=1 \text{ and }
Y_{ik}(z, g) \perp G_{ik} | R_{ik}=0$
%\[Y_{ik}(z, g) \perp Z_{ik} | R_{ik}=1 \qquad
%Y_{ik}(z, g) \perp G_{ik} | R_{ik}=0\]
\end{assumption}

\noindent Assumption \ref{ass:unc_ego} states that for index participants the treatment is randomized (it doesn't depend on the potential outcomes), and for network members the treatment of their indexes is randomized. 
Due to the randomization of the treatment to the index participants and under Assumption \ref{ass:nonoverlap}, given the participants' index status, Assumption \ref{ass:unc_ego} is satisfied. Here, Assumption \ref{ass:nonoverlap} ensures that $G_{ik}=0$ given $R_{ik}=1$, and  $Z_{ik}=0$ given
$R_{ik} = 0$, thus, Assumption \ref{ass:unc_ego} can also be written as $Y_{ik}(z, g) \perp Z_{ik} | (R_{ik}=1, G_{ik})$ and $Y_{ik}(z, g) \perp G_{ik} | (R_{ik}=0, Z_{ik})$.
Furthermore, we make here the assumption of random sampling.
\begin{assumption}[Random Sampling]
\label{ass:ran_sam}
$Y_{ik}(z, g) \perp R_{ik}$
\end{assumption}

\noindent This assumption states that potential outcomes do not depend on whether the unit is an index participant or a network member, that is, index participants are randomly sampled from the population and network members can also be seen as such.
Assumption \ref{ass:ran_sam} could be problematic if index participants self-select themselves to serve in this role. In this case, index participants and network members may differ in terms of their characteristics. 
%If this self-selection is a concern, further work should consider methods for adjustment for self-selection.
 If this self-selection is a concern,  one can estimate the individual effect only among index participants and the spillover effect only among network members. In this case, the comparison group for the estimation of the individual effect would only include untreated index participants, while the comparison group for the estimation of the spillover effect would only include network members of untreated index participants.

It can be shown that the unconfoundedness assumption (Assumption \ref{ass:unc_ego}), specific to the egocentric network-based design, and the random sampling assumption (Assumption \ref{ass:ran_sam}) guarantee the following unconfoundedness of the joint treatment. The proof of Assumption \ref{ass:unc} is given in Appendix \ref{Pf-A3}.
%of the Supplementary Materials. %\ref{Pf-A3}.

% To identify causal effects in the study sample, 
%Furthermore, we make the following assumption:
\begin{assumption}[Unconfoundedness of the joint treatment]
\label{ass:unc}
$Y_{ik}(z, g) \perp Z_{ik}, G_{ik}$
\end{assumption}
\noindent This assumption implies that both the individual treatment $Z_{ik}$ and the number of treated neighbors $G_{ik}$ are as good randomized.
Under Assumptions \ref{ass:nonoverlap}, \ref{ass:unc_ego} and \ref{ass:ran_sam}, then Assumption \ref{ass:unc} is satisfied for $(z,g)=\{(0,0), (0,1), (1,0)\}$. 
% It is worth noting that this assumption is satisfied by the egonetwork-based randomized design, under random sampling. 
% In our egonetwork-based randomized design, Assumption 3 guarantees for following unconfoundedness assumptions for the egonetwork-based design. 
%In fact, the 
%\noindent This assumption is required for the identification of causal effects in the sample.

Finally, to relate observed outcomes to potential outcomes, we make the following assumption, known as `consistency'.
\begin{assumption}[Consistency]
\label{ass:cons}
Denote by $Y_{ik}, Z_{ik}, G_{ik}$ the observed values of the outcome, the treatment, and the number of treated neighbors of unit $i$ in egonetwork $k$. Then, the following holds:
$Y_{ik}=Y_{ik}(Z_{ik}, G_{ik})$
\end{assumption}

\subsubsection{Proof of Assumption \ref{ass:unc} \label{Pf-A3}}
%\ref{ass:unc}

In this section, we show Assumption \ref{ass:unc} 
%\ref{ass:unc} 
is satisfied for $(Z_{ik},G_{ik})=\{ (0,0),(0,1),(1,0) \}$. We fist give the general form of $P\{Z_{ik}=z,G_{ik}=g|Y_{ik}(z',g')\}$:
\[
\begin{aligned}
P\{Z_{ik}=z,G_{ik}=g|Y_{ik}(z',g')\}& = \sum\limits^{1}_{r=0} P\{Z_{ik}=z,G_{ik}=g|Y_{ik}(z',g'),R_{ik}=r\}P(R_{ik}=r|Y_{ik}(z',g')\}\\
& = \sum\limits^{1}_{r=0} P\{Z_{ik}=z,G_{ik}=g|Y_{ik}(z',g'),R_{ik}=r\}P(R_{ik}=r)\\
& =  P\{Z_{ik}=z|Y_{ik}(z',g'),G_{ik}=g,R_{ik}=1\}P(G_{ik}=g|R_{ik}=1 )P(R_{ik}=1)\\
& \quad +  
P\{G_{ik}=g|Y_{ik}(z',g'),Z_{ik} = z, 
R_{ik}=0\}P(Z_{ik}=z|R_{ik}=0 )P(R_{ik}=0), \\
% & =  P(Z_{ik}=z|R_{ik}=1)P(R_{ik}=1) +  P\{G_{ik}=g|R_{ik}=0\}P(R_{ik}=0)\\
% & =  P(Z_{ik},G_{ik}|R_{ik}=1)P(R_{ik}=1) +  P\{Z_{ik},G_{ik}|R_{ik}=0\}P(R_{ik}=0)\\
% &=P(Z_{ik},G_{ik})
\end{aligned}
\]
where the second step is due to Assumption \ref{ass:ran_sam}. 
%\ref{ass:ran_sam}. 
Then we show Assumption \ref{ass:unc} 
%\ref{ass:unc} 
case by case.

\noindent \textit{1. Case for $(Z_{ik},G_{ik})=(0,0)$:}
\begin{align}
P\{Z_{ik}=0,G_{ik}=0|Y_{ik}(z',g')\} & =  P\{Z_{ik}=z|Y_{ik}(z',g'),G_{ik}=0,R_{ik}=1\}P(R_{ik}=1) \nonumber \\
&  +   P\{G_{ik}=g|Y_{ik}(z',g'),Z_{ik} = 0, R_{ik}=0\}P(R_{ik}=0) \nonumber  \\
 & =  P(Z_{ik}=0|G_{ik}=0,R_{ik}=1)P(R_{ik}=1) \nonumber  \\
 & +  P\{G_{ik}=0|Z_{ik} = 0,R_{ik}=0\}P(R_{ik}=0)\nonumber  \\
 & =  P(Z_{ik}=0,G_{ik}=0|R_{ik}=1)P(R_{ik}=1)  \nonumber  \\ 
 & +  P\{Z_{ik}=0,G_{ik}=0|R_{ik}=0\}P(R_{ik}=0) \nonumber \\
 &=P(Z_{ik}=0,G_{ik}=0), \nonumber 
\end{align}
where the second step is due to Assumption \ref{ass:nonoverlap} and  Assumption \ref{ass:unc_ego}.   %\ref{ass:nonoverlap} and Assumption \ref{ass:unc_ego}.

\noindent \textit{2. Case for $(Z_{ik},G_{ik})=(1,0)$:}
\begin{align}
P\{Z_{ik}=1,G_{ik}=0|Y_{ik}(z',g')\} & =  P\{Z_{ik}=1|Y_{ik}(z',g'),G_{ik}=0,R_{ik}=1\}P(R_{ik}=1) \nonumber \\
 & =  P(Z_{ik}=1|G_{ik}=0,R_{ik}=1)P(R_{ik}=1) \nonumber \\
 & =  P(Z_{ik}=1,G_{ik}=0|R_{ik}=1)P(R_{ik}=1)  \nonumber \\
  & =  P(Z_{ik}=1,G_{ik}=0|R_{ik}=1)P(R_{ik}=1)  \nonumber \\  & +P(Z_{ik}=1,G_{ik}=0|R_{ik}=0)P(R_{ik}=0)   \nonumber \\
 &=P(Z_{ik}=1,G_{ik}=0),\nonumber
\end{align}
where the first step is due to  Assumption \ref{ass:nonoverlap}, 
Assumption \ref{ass:unc_ego} 
and $P(Z_{ik}=1|R_{ik}=0 )=0$, the last step is due to $P(Z_{ik}=1,G_{ik}=0|R_{ik}=0)=0$.

\noindent \textit{3. Case for $(Z_{ik},G_{ik})=(0,1)$:} 
\begin{align}
P\{Z_{ik}=0,G_{ik}=1|Y_{ik}(z',g')\} & =  P\{G_{ik}=1|Y_{ik}(z',g'),Z_{ik}=0,R_{ik}=0\}P(R_{ik}=0) \nonumber \\
 & =  P(G_{ik}=1|Z_{ik}=0,R_{ik}=0)P(R_{ik}=0) \nonumber \\
 & =  P(Z_{ik}=0,G_{ik}=1|R_{ik}=0)P(R_{ik}=0) \nonumber \\
 &=P(Z_{ik}=0,G_{ik}=1), \nonumber
\end{align}
where the first step is due to  Assumption \ref{ass:nonoverlap}, 
Assumption \ref{ass:unc_ego}, 
%\ref{ass:unc_ego} 
and $P(G_{ik}=1|R_{ik}=1 )=0$, the last step isdue to $P(Z_{ik}=0,G_{ik}=G|R_{ik}=1)=0$.

Assumption \ref{ass:unc}  
%\ref{ass:unc} 
is satisfied by combing the results of Cases 1-3 above. 
%INTEGRATE THIS ABOVE

\clearpage

\subsection{Non-parametric Identification \label{identification}}

Given the characteristics of an egocentric network-based design, and under Assumptions \ref{ass:nonoverlap}, \ref{ass:neighint}, \ref{ass:unc}, and \ref{ass:cons}, 
we can identify the causal estimands from the observed data. In the following subsections, we report the identification results for each estimand and their proof.

\subsubsection{Non-parametric Identification of $\tau$ \label{iden-tau}}

\noindent For the individual effect $\tau$, we have
\[
\begin{aligned}
\tau=\E\big[Y_{ik}(1,0)-Y_{ik}(0,0)\big] = \E\{Y_{ik}| Z_{ik}=1 \}-\E\{Y_{ik}|Z_{ik}=0, G_{ik}=0\}.
\end{aligned}\]

\textbf{Proof:} We have
\[
\begin{aligned}
\E\big[Y_{ik}(1,0)\big] % &=  \E\{Y_{ik}(1,0)|R_{ik}=1\} \\
& =   \E\{Y_{ik}(1,0)|Z_{ik} =1, G_{ik} =0 \} \\
&=  \E\{Y_{ik}|Z_{ik} =1, G_{ik} =0 \} \\
%&=  \E\{Y_{ik}|Z_{ik} =1, R_{ik}=1 \} \\
&=  \E\{Y_{ik}|Z_{ik} =1\} ,
\end{aligned}
\]
% where the first step is because the index participant is randomly sample from the whole population (Assumption \ref{ass:ran_sam}), the second step is due to the unconfoudedness assumption (Assumption \ref{ass:unc}) and the third step is using the consistency property. 
where the first step is due to the unconfoudedness assumption (Assumption \ref{ass:unc}), 
%\ref{ass:unc}),  
the second step is using the consistency property (Assumption \ref{ass:cons}), 
%\ref{ass:cons}, 
and the third step is due to the egocentric randomization. We also have

We also have
\[
\begin{aligned}
\E\big[Y_{ik}(0,0)\big] %&=  \E\{Y_{ik}(0,0)|R_{ik}=0\}\\
& = \E\{Y_{ik}(0,0)|Z_{ik}=0, G_{ik}=0\}\\
%& = \E\{Y_{ik}|R_{ik}=0,Z_{ik}=0, G_{ik}=0\}\\
& = \E\{Y_{ik}|Z_{ik}=0, G_{ik}=0\},
\end{aligned}
\]
where the first step is due to the unconfoudedness assumption (Assumption \ref{ass:unc}), 
%\ref{ass:unc}),  
and the second step is using the consistency property (Assumption \ref{ass:cons}). %\ref{ass:cons}. 
As a result,
\[
\E\big[Y_{ik}(1,0)-Y_{ik}(0,0)\big] = \E\big[Y_{ik}(1,0)\big]-\E\big[Y_{ik}(0,0)\big]  =\E\{Y_{ik}| Z_{ik}=1 \}-\E\{Y_{ik}|Z_{ik}=0, G_{ik}=0\}.
\]

\subsubsection{Non-parametric Identification of $\delta$ \label{iden-delta}}
%\tcr{zhibing: checked}

\noindent For the spillover effect $\delta$, we have
\[
\begin{aligned}
  \delta=\E\big[Y_{ik}(0,1)-Y_{ik}(0,0)\big]=  \E[Y_{ik}| G_{ik}=1]-\E[Y_{ik}|Z_{ik}=0, G_{ik}=0].
\end{aligned}\]

\textbf{Proof:} Similar to the proof of identification for $\tau$, for $\delta$ we have
\[
\begin{aligned}
\E\big[Y_{ik}(0,1)\big] % &=  \E\{Y_{ik}(1,0)|R_{ik}=1\} \\
& =   \E\{Y_{ik}(0,1)|Z_{ik} =0, G_{ik} =1 \} \\
&=  \E\{Y_{ik}|Z_{ik} =0, G_{ik} =1 \} \\
%&=  \E\{Y_{ik}|Z_{ik} =1, R_{ik}=1 \} \\
&=  \E\{Y_{ik}|G_{ik} =1\} ,
\end{aligned}
\]
and 
\[
\begin{aligned}
\E\big[Y_{ik}(0,0)\big] % &=  \E\{Y_{ik}(1,0)|R_{ik}=1\} \\
& =   \E\{Y_{ik}(0,0)|Z_{ik} =0, G_{ik} =0 \} \\
&=  \E\{Y_{ik}|Z_{ik} =0, G_{ik} =0 \}. \\
%&=  \E\{Y_{ik}|Z_{ik} =1, R_{ik}=1 \} \\
%&=  \E\{Y_{ik}|Z_{ik} =0\} ,
\end{aligned}
\]
As a result, $\delta =  \E[Y_{ik}| G_{ik}=1]-\E[Y_{ik}|Z_{ik}=0, G_{ik}=0]$.

\subsubsection{Non-parametric Identification of  Overall effect $O$ \label{iden-o}}

\noindent For the overall effect $O$, we have
% \begin{equation}
% \begin{aligned}
%     O &=  E \left[ \frac{1}{n_k+1}\left(Y_{1k}(1,0) + \sum_{i=2}^{n_k+1}Y_{ik}(0,1)\right)\right]
%     % \left\{ \sum^{1}_{z=0} E \left[ Y_{ik}(z,1-z)\right]\cdot Pr(Z_{1k}=z)\right\} 
%     - \E[Y_{ik}(0,0)] \\
%     &= E  \left[ E \left\{ \frac{1}{n_k+1} Y_{1k}(1,0) -\frac{1}{n_k+1} Y_{ik}(0,0) \bigg\vert    k  \right\} \right] + E  \left[E \left\{  \frac{1}{n_k+1}\sum_{i=2}^{n_k+1}Y_{ik}(0,1) - \frac{n_k}{n_k+1}   Y_{ik}(0,0)\bigg\vert    k  \right\}\right]   \\
%     &=  E \left[\frac{1}{n_k+1}  E \left\{   Y_{1k}(1,0) - Y_{ik}(0,0)\right\}  \right] +  E \left[ \frac{1}{n_k+1} \sum_{i=2}^{n_k+1}E \left\{  Y_{ik}(0,1) -   Y_{ik}(0,0) \right\} \right]   \\
%         &=  E \left[ \frac{1}{n_k+1} \tau \right]+ E \left[\frac{n_k}{n_k+1}  E\left\{  Y_{ik}(0,1) -  Y_{ik}(0,0) \right\}  \right] \\
%   &=  E \left[ \frac{1}{n_k+1} \tau + \frac{n_k}{n_k+1} \delta  \right] \\
%     &=  E \left[ \frac{1}{n_k+1}\left(\tau + n_k\delta\right) \right]  .
%     \end{aligned}
%     \nonumber
% \end{equation}
\begin{equation*}
\begin{aligned}
\label{eq:overall_proof}
O&=  \E \left[\sum_{z=0}^1\sum_{g=0}^1 Y_{ik}(z,g)Pr(Z_{ik}=z, G_{ik}=g|Z_{1k}=1)\right]\\&\quad - \E \left[\sum_{z=0}^1\sum_{g=0}^1   (Y_{ik}(z,g)Pr(Z_{ik}=z, G_{ik}=g|Z_{1k}=0)\right]
\\&= \E \left[\sum_{r=0}^1\sum_{z=0}^1\sum_{g=0}^1 Y_{ik}(z,g)Pr(Z_{ik}=z, G_{ik}=g|Z_{1k}=1, R_{ik}=r)Pr(R_{ik}=r)\right]\\&\quad - \E \left[\sum_{r=0}^1\sum_{z=0}^1\sum_{g=0}^1   (Y_{ik}(z,g)Pr(Z_{ik}=z, G_{ik}=g|Z_{1k}=1, R_{ik}=r)Pr(R_{ik}=r)\right]
\\&
=\E \left[ Y_{ik}(1,0)Pr(R_{ik}=1)+ Y_{ik}(0,1)Pr(R_{ik}=0)\right]- \E \left[\sum_{r=0}^1 (Y_{ik}(0,0)Pr(R_{ik}=r)\right]
    \\&= \tau Pr(R_{ik}=1) +\delta Pr(R_{ik}=0),
    \end{aligned}
\end{equation*}
where we have used that $Pr(Z_{ik}=1, G_{ik}=0|Z_{1k}=1, R_{ik}=1)=1$, $Pr(Z_{ik}=0, G_{ik}=1|Z_{1k}=1, R_{ik}=0)=1$, and $Pr(Z_{ik}=0, G_{ik}=0|Z_{1k}=0, R_{ik}=r)=1$.
Then the identification of $O$ can be shown by plugging in the identification results of $\tau$ and $\delta$.

%\clearpage

\subsection{Variance for Regression Coefficients \label{Var-X}}

We have $\Sigma_I = \lim_{K \to \infty} \sigma^2_{Y} (\U_{Ik}/K )^{-1}$ and $\U_I =  \lim_{K \to \infty} \U_{Ik}/K$.
Then $\U_I$ can be written as (\citealt{Yang2020}):
\[
\U_{Ik} = c S_{Ik}+d T_{Ik}
\]
with
\[
 S_{Ik} = \sum\limits^{K}_{k=1}  \begin{bmatrix}
 n +1 &  Z_{1k} & n  G_{2k}\\
 Z_{1k}  &Z^2_{1k} &  0 \\
  n  G_{2k} &0 & n  G^2_{2k}
\end{bmatrix}
\]
and
\[
T_{Ik} = \sum\limits^{K}_{k=1}  \begin{bmatrix}
 (n +1)^2 &  (n +1)Z_{1k} & n(n +1)  G_{2k}\\
 (n +1)Z_{1k}  & Z^2_{1k} &  n Z_{1k}G_{2k} \\
n(n +1)  G_{2k}&  n Z_{1k}G_{2k} & n^2  G^2_{2k},
\end{bmatrix}
\]
where $c  = \frac{1}{1-\rho_{Y}}$ and $d = -\frac{\rho_{Y}}{(1-\rho_{Y})(1+n\rho_{Y})}$.
Then 
\[
S_I = \lim_{K \to \infty} \frac{1}{K} S_{Ik} =  \begin{bmatrix}
 n +1 &  p & n  p\\
p  & p &  0 \\
  n  p&0 & n p
\end{bmatrix}
\]
and 
\[
T_I = \lim_{K \to \infty} \frac{1}{K} T_{Ik} =  \begin{bmatrix}
 (n +1)^2 &  (n +1)p & n(n +1)  p\\
 (n +1)p  & p &  n p \\
n(n +1) p &  n p & n^2 p
\end{bmatrix}.
\]
As a result,

\[
\U_I = cS_I + d T_I =\left[\begin{array}{c|cc}
 \{c+(n+1)d\} (n+1)  &  \{c+(n+1)d\} p & \{c+(n+1)d\}  n p \\
 \hline
 \{c+(n+1)d\} p & (c+d) p & n p d \\
 \{c+(n+1)d\}  n p  & n p d & (c+nd) n p \\
\end{array} \right]
\]

\subsection{Derivation of $\Sigma_{\tau\delta}$ \label{Sig-1}}

From Section \ref{Var-X}, we have
\[
\U_I = \left[\begin{array}{c|cc}
 \{c+(n+1)d\} (n+1)  &  \{c+(n+1)d\} p & \{c+(n+1)d\}  np\\
 \hline
 \{c+(n+1)d\} p & (c+d)p & n p d \\
 \{c+(n+1)d\}  n p & n p d & (c+nd) n p \\
\end{array} \right]=\left[\begin{array}{c|c}
A  & B \\
 \hline
 C & D \\
\end{array} \right]
\]
with $c  = \frac{1}{1-\rho_{Y}}$ and $d = -\frac{\rho_{Y}}{(1-\rho_{Y})(1+n\rho_{Y})}$.  We split $U_I$ into a $2\times2$ blockwise matrix for further matrix inversion.
$\Sigma_{\tau\delta}$, the lower-right element of $\Sigma_I$, which is $\lim_{K \to \infty} \sigma^2_{Y} (\U_{Ik}/K )^{-1}$,  is the corresponding covariance for $\tau$ and $\delta$, is calculated by $\sigma^2_{Y} \times (D-CA^{-1}B)^{-1}$.
Thus, $\Sigma_{\tau\delta}$ can be written as
\[
\Sigma_{\tau\delta} =\frac{\sigma^2_{Y}}{E_4} \begin{bmatrix}
 A_4 & B_4 \\
 C_4 & D_4\\
\end{bmatrix},
\]
\[
A_4 =  (c+nd) n p - \frac{n^2 p^2}{n+1} \{ c+(n+1)d \}=ncp \left\{1-\frac{np}{(n+1)}\right\}+n^2 d p (1-p) ,
\]
\[
B_4 = C_4 = \frac{n p^2}{n+1} \{ c+(n+1)d \}-n d p =\frac{ncp^2}{n+1} - ndp(1-p),
\]
\[
D_4 = (c+d) p -\frac{p^2}{n+1}\{ c+(n+1)d \}=cp \left\{1-\frac{p}{(n+1)}\right\}+ d p (1-p) ,
\]
and
\[
E_4 =A_4 \times D_4-B_4\times C_4
%n p^2 \left[  c^2 +cd+ncd - \frac{c+d+cp+ndp^2+2dp}{n+1}\{ c+(n+1)d \} +\frac{p-np^2}{(n+1)^2}\{ c+(n+1)d \}^2 \right]
\]
We let $\sigma^2_Z = p(1-p)$, $m_1 = p(1-\frac{p}{n+1})$ and $m_2 =p(1-\frac{np}{n+1}) $. Then
\[
A_4 = nc m_2+n^2d\sigma_Z^2, \quad B_4=C_4 = nc(p-m_1)-nd\sigma^2_Z, \quad D_4 = cm_1+d\sigma^2_Z
\]
and
\[
\begin{aligned}
E_4 & =  nc\{ cp(m_1+m_2-p)+d \sigma^2_Zp(1+n)  \} \\
    & = ncp\{  cp(1-p)+d\sigma^2_Z (1+n) \} \\
    & = ncp\{  c\sigma^2_Z+d\sigma^2_Z (1+n) \} \\
    & = ncp\sigma^2_Z\{c+d (1+n)\}.
\end{aligned}
\]
Using the fact that $(p-m_2) =n(p-m_1) $, the formula of $\Sigma_{\tau\delta}$ can be simplified as 
\[
\Sigma_{\tau\delta} =\sigma^2_{Y} \begin{bmatrix}
\frac{c m_2 +nd\sigma^2_Z}{cp\sigma^2_Z\{c+d(1+n)\}} &  \frac{c(p-m_1) - d\sigma^2_Z}{cp\sigma^2_Z\{c+d(1+n)\}}\\
\frac{c(p-m_1) - d\sigma^2_Z}{cp\sigma^2_Z\{c+d(1+n)\}} & \frac{cm_1+d\sigma^2_Z}{ncp\sigma^2_Z\{c+d(1+n)\}} \\
\end{bmatrix}.
\]

\clearpage

\subsection{The Minimum Number of Network Required for HISpJ \label{N-HISpJ} }

In this section, we provide the derivation of the minimum number of index participants required for HISpJ. Our derivation is based on a general jointly hypothesis test $H_0: L \btheta_{J} = 0$, where $\btheta_{J}=(\tau, \delta)'$ and $L$ is the matrix to construct the jointly hypothesis test. Then the Wald test statistic is given by 
\[
Q_{J} = K(L\hat{\btheta}_J)'(L\hat{\Sigma}^{-1}_{\tau\delta}L')(L\hat{\btheta}_J).
\]

We have known that from Generalized least squares, 
 $\sqrt{K}(\hat{\btheta}_J - \btheta_J) \xrightarrow{d} N(0, \Sigma_{\tau\delta})$ as $K \to \infty$. Then 
$\sqrt{K}(L\hat{\btheta}_J - L\btheta_J) \xrightarrow{d} N(0, L \Sigma_{\tau\delta} L')$ as $K \to \infty$. As a result, $L\hat{\btheta}_{J}$ can be presented as
\[
L\hat{\btheta}_{J} = \frac{1}{\sqrt{K}} (L \Sigma_{\tau\delta}L')^{\frac{1}{2}}W_q+L\btheta_J,
\]
where $W_q$ is a $q$-length vector following a standard multivariate normal distribution.
Then we can write the test statistic, $Q_{J}$ as 
\[
Q_{J} =  K\left\{\frac{1}{\sqrt{K}}( L \Sigma_{\tau\delta}L')^{\frac{1}{2}}W_q+L\btheta_J \right\}'(L\hat{\Sigma}^{-1}_{\tau\delta}L')\left\{\frac{1}{\sqrt{K}} (L \Sigma_{\tau\delta}L')^{\frac{1}{2}}W_q+L\btheta_J\right\}.
\]
%By Slutsky's theorem, 
It has the same asymptotic distribution as
\[
\begin{aligned}
Q^*_{J} & = \{ (L \Sigma_{\tau\delta}L')^{\frac{1}{2}}W_q+\sqrt{K}L\btheta_J\}'(L\Sigma^{-1}_{I4}L')\{ (L \Sigma_{\tau\delta}L')^{\frac{1}{2}}W_q+\sqrt{K}L\btheta_J\} \\
& = \{ W_q+\sqrt{K}(L\Sigma_{\tau\delta}L')^{-\frac{1}{2}}L\btheta_J\}' \{ W_q+\sqrt{K}(L\Sigma_{\tau\delta}L')^{-\frac{1}{2}}L\btheta_J\} \\
& = \sum\limits^{q}_{l=1} (W_{q,l}+\sqrt{K}L_{\btheta,l})^2,
\end{aligned}
\]
where $W_{q,l}$ is the $l$th element in $W_q$ and $L_{\btheta,l}$ is the $l$th element of $L_{\btheta} = (L\Sigma_{\tau\delta}L')^{-\frac{1}{2}}L\btheta_J$.

Then the minimum number of networks required for HISpJ to have power $\pi$ should satisfy
\begin{equation}
 \Pr\{Q^*_{J} \geq \chi^2_{1-\alpha}(q) |L\btheta_J = L\Delta_J   \} \geq \pi. \label{Q-J} %\tag{S.1}
\end{equation}

So it suffices to let
\[
K \sum\limits^{q}_{l=1} L^2_{\btheta,l} \geq \upsilon(\chi^2_{1-\alpha}(q),\pi,q)
\]
to satisfy \eqref{Q-J}.
Thus, we have that
\[
\begin{aligned}
K & \geq \frac{\upsilon(\chi^2_{1-\alpha}(q),\pi,q)}{\sum\limits^{q}_{l=1} L^2_{\btheta,l} } \\
& = \frac{\upsilon(\chi^2_{1-\alpha}(q),\pi,q)}{ (L\btheta_{J})(L\Sigma_{I4}L')^{-1}(L\btheta_J)'}.
\end{aligned}
\]

To be specific, for the hypothesis test of testing individual and spillover effects simultaneously, we use $\upsilon(\chi^2_{1-\alpha}(2),\pi,2)$ and  $L$ is the $2 \times 2$ diagonal matrix to find $K_{J}$.

\clearpage

\subsection{The first derivative of $K$ with respect to $\rho_Y$, $p$, $n$, and effect size \label{DRV-K}}

 In this section, We provide the first derivative of $K$ with respect to $\rho_Y$, $p$, $n$, and effect size, respectively, to investigate how the number of index participants changes with these design parameters.

\subsubsection{The first derivative of $K_{\tau}$\label{DRV-K-T}}

 We first provide the first derivative of $K_{\tau}$ in equation (\ref{K-tau}) with respect to $\rho_{Y}$:
\begin{equation}
\begin{aligned}
\frac{\partial K_{\tau}}{\partial  \rho_Y } &=\frac{\sigma^2_Y n (z_{1-\alpha/2}+z_{\pi})^2}{\Delta_{\tau}^2\cdot(n+1)(1-p)}.
\end{aligned} \label{dr-t-r} % \tag{S.2}
\end{equation}
We can learn that \eqref{dr-t-r} is always positive, so $K_{\tau}$ will monotonically increase as $\rho_Y$ increases.

The first derivative of $K_{\tau}$ with respect to $n$:
\begin{equation}
\begin{aligned}
\frac{\partial K_{\tau}}{\partial  n } =  -\frac{ \sigma^2_Y (z_{1-\alpha/2}+z_{\pi})^2 \cdot\left(\rho_Y-1\right)}{\Delta_{\tau}^2\cdot\left(p-1\right)\left(n+1\right)^2} 
\end{aligned}\label{dr-t-n}%  \tag{S.3}
\end{equation}
We can observe that $K_{\tau}$ will monotonically decrease as $n$ increases since \eqref{dr-t-n} is always negative.

The first derivative of $K_{\tau}$ with respect to $p$:
\begin{equation}
\begin{aligned}
\frac{\partial K_{\tau}}{\partial  p } &=\frac{\sigma^2_Y (z_{1-\alpha/2}+z_{\pi})^2\cdot\left\{\left(\rho_Y-1\right)np^2+\left(2n+2\right)p-n-1\right\}}{\Delta_{\tau}^2\cdot\left(n+1\right)\left(p-1\right)^2p^2} 
\end{aligned}\label{dr-t-p} %\tag{S.4} 
\end{equation}
From \eqref{dr-t-p}, we cannot observe a monotone change of $K_{\tau}$ with $p$. However, given other parameters fixed, by letting $\frac{\partial K_{\tau}}{\partial  p } =0$ and solve it for $p$, we could obtain the unique solution of $p \in [0,1]$:
\begin{equation}
\begin{aligned}
p = \frac{(n+1)+\sqrt{(n+1)(1+n\rho_Y)}}{(1-\rho_Y)n} \nonumber
\end{aligned},
\end{equation}
which minimize $K_{\tau}$. 

The first derivative of $K_{\tau}$ with respect to $\Delta_{\tau}$ is
\begin{equation}
\begin{aligned}
\frac{\partial K_{\tau}}{\partial  \Delta_{\tau} } &=\frac{2 \sigma^2_Y (z_{1-\alpha/2}+z_{\pi})^2 \cdot\left\{\left(\rho_Y-1\right)np+n+1\right\}}{\left(n+1\right)\left(p-1\right)p \Delta_{\tau}^3}\end{aligned}\label{dr-t-d} %\tag{S.5}
\end{equation}
where we can observe that  $K_{\tau}$ will monotonically decreases as $\Delta_{\tau}$ increases.

\subsubsection{The first derivative of $K_{\delta}$ \label{DRV-K-D}}

We first provide the first derivative of $K_{\delta}$ in (\ref{K-delta}) with respect to $\rho_{Y}$:
\begin{equation}
\begin{aligned}
\frac{\partial K_{\delta}}{\partial  \rho_Y } &=\dfrac{\sigma^2_Y (z_{1-\alpha/2}+z_{\pi})^2\cdot\left(p+n^2-1\right)}{\Delta_{\delta}^2n\cdot\left(n+1\right)\left(1-p\right)p} 
\end{aligned} \label{dr-d-r}% \tag{S.6}
\end{equation}
We can learn that \eqref{dr-d-r} is always positive, so $K_{\delta}$ will monotonically increase as $\rho_Y$ increases.

The first derivative of $K_{\delta}$ with respect to $n$:
\begin{equation}
\begin{aligned}
\frac{\partial K_{\delta}}{\partial  n} &=\frac{ \sigma^2_Y (z_{1-\alpha/2}+z_{\pi})^2\cdot\left(\rho_Y-1\right)\cdot\left(n^2+\left(2-2p\right)n-p+1\right)}{\Delta_{\delta}^2\cdot\left(1-p\right)p n^2\cdot\left(n+1\right)^2} 
\end{aligned}\label{dr-d-n} %\tag{S.7}
\end{equation}
We can learn that \eqref{dr-d-n} is always negative, so $K_{\delta}$ will monotonically decrease as $n$ increases.

The first derivative of $K_{\delta}$ with respect to $p$:
\begin{equation}
\begin{aligned}
\frac{\partial K_{\delta}}{\partial  p } &=\frac{ \sigma^2_Y (z_{1-\alpha/2}+z_{\pi})^2\cdot\left\{\left(\rho_Y-1\right)p^2+\left(2\rho_Y n^2+2n-2\rho_Y +2\right)p-\rho_Y n^2-n+\rho_Y-1\right\}}{\Delta_{\delta}^2n\cdot\left(n+1\right)\left(p-1\right)^2p^2} \\
& = \frac{ \sigma^2_Y (z_{1-\alpha/2}+z_{\pi})^2\cdot\left[(\rho_Y-1)p^2 +(2p-1)\{\rho_Y(n-1)+1\}(n+1)\right]}{\Delta_{\delta}^2n\cdot\left(n+1\right)\left(p-1\right)^2p^2}
\end{aligned}\label{dr-d-p}%   \tag{S.8}
\end{equation}
From \eqref{dr-d-p}, we cannot observe a monotone change of $K_{\delta}$ with $p$. Given $p \in [0,1]$, there exist only one solution for $\frac{\partial K_{\delta}}{\partial  p } =0$, which is 
\begin{equation}
\begin{aligned}
p = \frac{(n^2\rho_Y+n-\rho_Y+1)\left(1-\sqrt{\frac{n^2\rho_Y+n-\rho_Y+1}{n^2\rho_Y+n-2\rho_Y+2}}\right)}{1-\rho_Y} \nonumber
\end{aligned}.
\end{equation}

%and when $p=0$, $\frac{\partial K_{\delta}}{\partial  p } <0$ and when $p=1$, $\frac{\partial K_{\delta}}{\partial  p }>0 $. %But when $p=0.5$, \eqref{dr-d-p} takes a minimal value given any other parameters.

The first derivative of $K_{\delta}$ with respect to $\Delta_{\delta}$ is
\begin{equation}
\begin{aligned}
\frac{\partial K_{\delta}}{\partial  \Delta_{\delta} } &=\frac{2\sigma^2_Y (z_{1-\alpha/2}+z_{\pi})^2\cdot\left\{\left(\rho_Y-1\right)p+\rho_Y n^2+n-\rho_Y+1\right\}}{n\cdot\left(n+1\right)\left(p-1\right)p\Delta_{\delta}^3}
\end{aligned} \label{dr-d-d} %\tag{S.9}
\end{equation}
where we can observe that $K_{\delta}$ will monotonically decrease as $\Delta_{\delta}$ increases.

\subsubsection{The first derivative of $K_{o}$ \label{DRV-K-O}}

We first provide the first derivative of $K_{o}$ in equation (\ref{K-o}) with respect to $\rho_{Y}$:
\begin{equation}
\begin{aligned}
\frac{\partial K_{o}}{\partial  \rho_Y } &=\frac{\sigma^2_Y (z_{1-\alpha/2}+z_{\pi})^2n}{\Delta_{o}^2\left(n+1\right)\left(1-p\right)p} 
\end{aligned} \label{dr-o-r} %\tag{S.10}
\end{equation}
We can learn that \eqref{dr-o-r} is always positive, so $K_{o}$ will monotonically increase as $\rho_Y$ increases.

The first derivative of $K_{o}$ with respect to $n$:
\begin{equation}
\begin{aligned}
\frac{\partial K_{o}}{\partial  n} &=
\dfrac{\sigma^2_Y (z_{1-\alpha/2}+z_{\pi})^2\cdot\left[\left\{\left(\rho_Y+1\right)\Delta_{\delta}-2\rho_Y \Delta_{\tau}\right\}n+2\Delta_{\delta}+\left(-\rho_Y-1\right)\Delta_{\tau}\right]}{\left(p-1\right)p\cdot\left(\Delta_{\delta}+\Delta_{\tau}\right)^3}
\end{aligned} \label{dr-o-n} % \tag{S.11}
\end{equation}
We observe that the change of $K_{O}$ is related to the magnitude of the effect size of $\tau$ and $\delta$.

The first derivative of $K_{o}$ with respect to $p$:
\begin{equation}
\begin{aligned}
\frac{\partial K_{o}}{\partial  p} &=
\frac{\sigma^2_Y (z_{1-\alpha/2}+z_{\pi})^2 \cdot\left(\rho_Y n+1\right)\cdot\left(2p-1\right)}{\Delta_{\delta}^2\cdot\left(n+1\right)\left(p-1\right)^2p^2}
\end{aligned} \label{dr-o-p} % \tag{S.12}
\end{equation}
From \eqref{dr-o-p}, we can observe that when $p=0.5$, $K_{O}$ will take the optimal value. Meanwhile, $K_{O}$ decreases first and then increases as $p$ increases from 0 to 1.

\subsubsection{The first derivative of $K_{J}$ \label{DRV-K-TD}}

We first provide the first derivative of $K_{J}$ in equation (\ref{K-td}) with respect to $\rho_{Y}$:
\begin{equation}
\begin{aligned}
\frac{\partial K_{J}}{\partial  \rho_Y } &=\frac{\sigma^2_Y \upsilon(\chi^2_{1-\alpha}(2),\pi,2) n}{\left(\Delta_{\delta}^2n+\Delta_{\tau}^2\right)\left(1-p\right)p} 
\end{aligned}\label{dr-td-r} % \tag{S.13}
\end{equation}
We can learn that \eqref{dr-td-r} is always positive, so $K_{J}$ will monotonically increase as $\rho_Y$ increases.

We first provide the first derivative of $K_{J}$ with respect to $n$:
\begin{equation}
\begin{aligned}
\frac{\partial K_{J}}{\partial  n} &=
\frac{\sigma^2_Y \upsilon(\chi^2_{1-\alpha}(2),\pi,2)\cdot\left(\Delta_{\delta}^2-\rho_Y \Delta_{\tau}^2\right)}{\left(p-1\right)p\cdot\left(\Delta_{\delta}^2n+\Delta_{\tau}^2\right)^2}
\end{aligned} \label{dr-td-n} %  \tag{S.14}
\end{equation}
We observe that the change of $K_{J}$ is related to the magnitude of the effect size of $\tau$ and $\delta$.

The first derivative of $K_{J}$ with respect to $p$:
\begin{equation}
\begin{aligned}
\frac{\partial K_{J}}{\partial  p} &=
\dfrac{\sigma^2_Y \upsilon(\chi^2_{1-\alpha}(2),\pi,2)\cdot\left(\rho_Y n+1\right)\cdot\left(2p-1\right)}{\left(\Delta_{\delta}^2n+\Delta_{\tau}^2\right)\left(p-1\right)^2 p^2}
\end{aligned} \label{dr-td-p} % \tag{S.15}
\end{equation}
From \eqref{dr-td-p}, we can observe that when $p=0.5$, $K_{J}$ will take the optimal value. Meanwhile, $K_{J}$ decreases first and then increases as $p$ increases from 0 to 1.

\subsection{Calculation of Network Size $n$ and the Minimum Detectable Effect Size (MDE)\label{nMDE}}

In the study design, we are also interested in calculating the required number of network members given a specified number of index participants, a hypothesized effect size, a desired power and Type I error, as well as the minimum detectable effect size (MDE) given a specified number of index participants and network members, a desired power and Type I error.

%for tau
Using \eqref{K-tau}, \eqref{K-delta} and \eqref{K-o},
the MDE of $\tau$, $\delta$ and $O$ are
\begin{equation}
\begin{aligned}
\Delta_{\tau} &= \sqrt{ \frac{\sigma^2_{Y}[n\{p (\rho_Y -1)+1\}+1] (z_{1-\alpha/2}+z_{\pi})^2}{(n+1)\sigma^2_Z K}}, \nonumber
\end{aligned}
\end{equation}
\begin{equation}
\Delta_{\delta}=\sqrt{ \frac{\sigma^2_{Y}\{(1-p)(1-\rho_{Y})+n(1+n\rho_{Y}) \} (z_{1-\alpha/2}+z_{\pi})^2}{n(n+1)\sigma^2_Z  K_{\delta}}}, \nonumber
\end{equation}
and
\begin{equation}
\Delta_{o}=\sqrt{ \frac{\sigma^2_{Y}(1+n\rho_Y) (z_{1-\alpha/2}+z_{\pi})^2}{(n+1)\sigma^2_Z K_o}}.
\nonumber
\end{equation}
The MDE of HISpJ incorporates $\Delta_{\tau}$ and $\Delta_{\delta}$ at the same time. Given the given $K$, $n$, $\rho_Y$, $\sigma^2_Y$, $p$ and fixed power and type I error rate, $\alpha$, we have the solution to the implicit equations 
\begin{equation}
 \Delta_{\tau}=\sqrt{\frac{\upsilon(\chi^2_{1-\alpha}(2),\pi,2) \sigma^2_{Y}(1+n\rho_Y)}{K_{J} \sigma^2_{Z}}-n\Delta_{\delta}^2}. \nonumber
\end{equation}
and 
\begin{equation}
\Delta_{\delta}=\sqrt{\frac{\upsilon(\chi^2_{1-\alpha}(2),\pi,2) \sigma^2_{Y}(1+n\rho_Y)}{nK_{J} \sigma^2_{Z}}-\frac{\Delta_{\tau} }{n}}. \nonumber
\end{equation}

\begin{figure}
\centerline{%
\includegraphics[width=\linewidth]{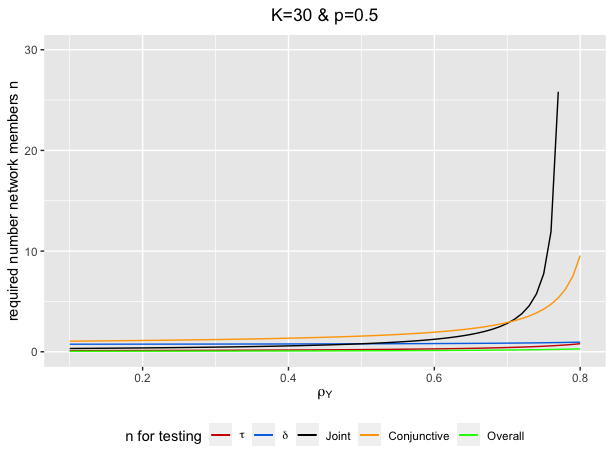}}
%% to include a figure, or to leave a blank space
\caption{Required number of network members for HIE, HSpE, HISpJ, HISpC, and HOE given different values of $\rho_Y \in [0.1,0.8] $ to achieve 80\% power: $\Delta_{\tau}=\Delta_{\delta}$=1, $\sigma^2_{Y} =1$, $p=0.5$, $K=30$.}
\label{fig:plot_n} %\tag{Figure S.1}
\end{figure}

To obtain the minimum number of network members for each index participant, we solve the sample size equation \eqref{K-tau}, \eqref{K-delta}, \eqref{K-td} and \eqref{K-o} 
with respect to $n$, given pre-specified $K$ and other sample size parameters for each hypothesis test. Then, the problem now involves finding the root of non-linear equations, which some packages in the statistical software could solve. In this paper, we use \textit{multiroot} in \textit{R} to obtain the solutions. To be specific, here we use AIE as an example to show the calculation details of $n$. From Section \ref{Cal_K}, 
we know the sample size equation of testing AIE is \eqref{K-tau}. 
Then the equation needs to be solved is $K_{\tau}-K=0$, where $K$ is the pre-specified sample size given required power and Type I error, $\alpha$. In most of the cases, we could obtain one or multiple solutions, where the smallest one is selected as the optimal result. However, under some situations, there is no solution for $n$ for each index participant given the sample size equation, and other design parameters. This phenomenon has been seen in other clustered studies. To overcome this obstacle, one could consider modifying the pre-specified parameters. Usually, we could increase the number of index participants, $K_{\tau}$, by each unit, and then solve the equation again until the solution is founded.
For HISpC, without the closed form of the sample size $K_C$, we used the power function \eqref{K-HISpC}
instead of the sample size equation to find $n$. We solve $\pi_C- \pi=0$, given a pre-specified power and other design parameters. Figure \ref{fig:plot_n} 
shows an example that how $n$ changes with $\rho_Y$ for HIE, HSpE, HOE, HISpJ, and HISpC by solving \eqref{K-tau}, \eqref{K-delta}, \eqref{K-td} and \eqref{K-o}
equal to 30 or \eqref{K-HISpC} 
equals to 80\%, given $p=0.5$, $\Delta_{\tau} = \Delta_{\delta}$, $\sigma^2_Y=1$. We could observe that for all five tests, $n$ increases as $\rho_Y$ increased. To note, for HISpJ, we cannot obtain a reasonable $n$ when $\rho_Y >0.77$. Under this situation, we could increase $K$ to overcome this problem, such as $n =0.85$ when $K=35$ and $\rho_Y=0.8$.

Solving the sample size equation or power function is not the only way to find $n$. One could treat this problem as an optimization problem by setting the power function as the objective function. Then, we could find the minimum $n$ maximizing the power with different parameter constraints, such as the test should at least achieve the required power, or the study has the maximum number of networks. 

%Under some situations, 

%\begin{equation}
%\begin{aligned}
%n_{{\tau}}& = \frac{K_{\tau}\sigma^2_Z \Delta_{\tau}^2-\sigma^2_{Y}(z_{1-\alpha/2}+z_{1-\lambda})^2}{(z_{1-\alpha/2}+z_{1-\lambda})^2\{p (\rho_Y -1)+1\}-K_{\tau}\sigma^2_Z \Delta_{\tau}^2}. \nonumber
%\end{aligned}
%\end{equation}
%\textcolor{blue}{We could notice that $n_{\tau}$ could take valid value at some combination of other sample size parameters}.

We aslo use HPTN 037 (\citealt{Latkin2009}) to demonstrate how
to estimate MDE and $n$ given $K$, pre-specified power and Type I error
for the hypothesis tests considered in this paper
We inverted equations \eqref{K-tau}, \eqref{K-delta} and \eqref{K-o}
to obtain the absolute minimum detectable effect sizes for the AIE, ASpE, and overall effect to be $\vert 0.34\vert$, $\vert 0.28\vert$, an $\vert 0.26 \vert$, respectively, for different value of $\rho_Y$ and $p$ (Table \ref{tab:2}), with pre-specified power $80\%$ and Type I error rate $5\%$. In Table \ref{tab:3}, we also provide the minimum required network sizes $n$ for the HIE, HSpE, HISpJ, HISpC, and HOE. Given $\tau=-0.35$, we cannot obtain $n_{\tau}$ and $n_{C}$ to achieve the required 80\% power and $\alpha=0.05$ for HIE when $p=0.30$. As we discussed in this section, we could increase $K$ to obtain an applicable $n$. For example, when we increase $K=250$, $n_{\tau} = 2.02$, and $n_C = 2.89$ for the case of $\rho_Y=0.10$ and $p=0.30$. 

\begin{table}[htbp]
  \centering
  \caption{Minimum absolute values of detectable effect sizes given 186 networks, $n=2$, $\sigma^2_Y=1$, $\alpha=0.05$ and 80\% power}  \label{tab:2}%
    \begin{tabular}{ccccc}
   \toprule
    \multicolumn{2}{c}{Design Parameters} & HIE   & HSPE  & HOE \\
     \hline
    $\rho_Y$ & p     & $|\Delta_{\tau}|$ & $|\Delta_{\delta}|$ &  $|\Delta_{o}|$ \\
  \hline
    \multirow{3}[2]{*}{0.10} & 0.50   & 0.34  & 0.28  & 0.26 \\
          & 0.30   & 0.41  & 0.32  & 0.28 \\
          & 0.70   & 0.34  & 0.30   & 0.28 \\
 \hline
    \multirow{3}[2]{*}{0.20} & 0.50   & 0.35  & 0.30   & 0.28 \\
          & 0.30   & 0.41  & 0.34  & 0.31 \\
          & 0.70   & 0.36  & 0.32  & 0.31 \\
  \hline
    \multirow{3}[2]{*}{0.05} & 0.50   & 0.40   & 0.28  & 0.25 \\
          & 0.30   & 0.34  & 0.31  & 0.27 \\
          & 0.70   & 0.34  & 0.29  & 0.27 \\
 \hline
    \end{tabular}%
\end{table}%

\begin{table}[htbp]
  \centering
 \caption{Minimum required network members given 186 index participants with $\tau = -0.35$, \\$\delta=-0.35$, $\sigma^2_Y=1$, $\alpha=0.05$ and 80\% power}  \label{tab:3}%
    \begin{tabular}{ccccccc}
 \toprule
    \multicolumn{2}{c}{Design Parameters} & \multicolumn{1}{l}{HIE} & \multicolumn{1}{l}{HSPE} & \multicolumn{1}{l}{HISpJ} & \multicolumn{1}{l}{HISpC} & \multicolumn{1}{l}{HOE} \\
   \hline
    $\rho_Y$ & p     & $n_{\tau}$ & $n_{\delta}$ & $n_{J}$ & $n_{C}$& $n_{o}$ \\
   \hline
    \multirow{3}[2]{*}{0.10} & 0.50   & 1.62  & 1.1   & 0.84 & 2.40 & 0.45 \\
          & 0.30   & ND    & 1.57  & 1.28  &  ND & 0.78 \\
          & 0.70   & 1.67  & 1.22  & 1.28 &  2.30 & 0.78 \\
    \hline
    \multirow{3}[2]{*}{0.20} & 0.50   & 2.28  & 1.18  & 1.06 & 3.29 & 0.53 \\
          & 0.30   & ND    & 1.75  & 1.72 & ND & 0.97 \\
          & 0.70   & 2.37  & 1.39  & 1.72 &  3.23 & 0.97 \\
   \hline
    \multirow{3}[2]{*}{0.05} & 0.50   & 1.41  & 1.07  & 0.77 &  2.14 & 0.41 \\
          & 0.30   & ND    & 1.50   & 1.14 & ND & 0.71 \\
          & 0.70   & 1.45  & 1.15  & 1.14  & 2.03  & 0.71 \\
  \hline
     \multicolumn{6}{p{10cm}}{In this example, ND represents for not discovered.}
    \end{tabular}%
\end{table}%

\clearpage

\subsection{Alternative Regession-based Estimators}

\label{AlterModel}

\subsubsection{Statistical Model \label{AlterModel-2}}

We consider two separate regression models for estimating the AIE and the ASpE based on the identification results in Section \ref{Causal-Est}. Implicitly, the AIE is estimated by comparing the outcomes of index participants in the networks with intervention with those in the networks without intervention, while the ASpE is estimated by comparing the outcomes between networks members in networks with/out intervention (Figure \ref{fig:design}). For $i=1$, we have
\begin{equation}
Y_{1k}=\gamma_{\tau} + \tau Z_{1k} + \epsilon_{1k}  \label{m2-tau} %\tag{S.16}
\end{equation}
and for $i=2,...,(n+1)$,
\begin{equation}
Y_{ik} = \gamma_{\delta} +\delta G_{ik} + u_k + \epsilon_{ik}, \label{m2-delta} %\tag{S.17}
\end{equation}
where we assume $\epsilon_{1k} \sim N(0,\sigma^2_{e1}) $, $\epsilon_{ik} \sim N(0,\sigma^2_e) $, $u_k \sim N(0,\sigma^2_u)$ and $\epsilon_{ik} \perp u_k$. We also assume $\sigma^2_e$ and $\sigma^2_u$ are known. We let $\theta_I = (\gamma_{\tau}, \tau)'$ and  $\theta_{II} = (\gamma_{\delta}, \delta)'$.

In model \eqref{m2-tau}, $\gamma_{\tau}$ is the mean of index participants in networks without intervention, which is estimated by $\E[Y_{1k}|Z_{1k}=0]$. The estimate of $\tau$ in model \eqref{m2-tau} is $\E[Y_{1k}|Z_{1k}=1]-\E[Y_{1k}|Z_{1k}=0]$, which is same as the identification of the AIE using the observed data in Section \ref{Causal-Est}. In model \eqref{m2-delta}, $\gamma_{\delta}$ is the mean of network members in the networks without intervention, where can be estimated by $\E[Y_{ik}|G_{ik}=0]$. Meanwhile, $\delta$ can estimated through $\E[Y_{ik}|G_{ik}=1]-\E[Y_{ik}|G_{ik}=0]$, which is same as the identification of the ASpE in Section  \ref{Causal-Est}.

To proceed, we first show the estimator of $\theta_I$ using model \eqref{m2-tau}. We let $\Y_1 = (Y_{11},Y_{12},...,Y_{1K})'$, $\Z_1 = ((1, Z_{11})',..., (1, Z_{12})',...,(1, Z_{1K})')'$. The total variance of $Y_{1k}$, denoted as $\sigma^2_{Y_1}$, is given by $\sigma^2_{Y_1} =\sigma_{e1}^2 $. For any $k \neq k'$, $Y_{1k}$ and $Y_{1k'}$ are independent with each other. Then we know that the best linear unbiased estimator (BLUE) of $\beta_I $ is given by least square estimator:
\[
\hat{\theta}_I = (\Z_1' \Z_1)^{-1} \cdot \Z_1' \Y_1
\]
with
\[
\Var(\hat{\theta}_I) = \sigma^2_{\Y_1}U^{-1}_{Dk}, 
\]
where $U_{Dk} = \Z_1' \Z_1$. When $K \to \infty$, $\sqrt{K}(\hat{\beta}_I-\beta_I)$ is asymptotic normal distributed as $N(0, \Sigma_D)$, where 
\[
\Sigma_D = \lim_{K \to \infty} \sigma^2_{Y_1} (U_{Dk}/K )^{-1} = \frac{\sigma^2_{Y_1}}{p(1-p)}\begin{bmatrix}
p & -p \\
-p & 1
\end{bmatrix}.
\]

In model \eqref{m2-delta}, we let $\Y_{Sk} = (Y_{2k},...,Y_{(n+1)k})'$ and $\mbG_{Sk} = ((1,G_{2k})',...,(1,G_{(n+1)k})')'$. For $i > 1$, the total variance of $Y_{ik}$, denoted by $\sigma^2_{Y}$, is given by $\sigma^2_e +\sigma^2_u$, and for any $i \neq i'$, the covariance between to network members in same network is $\Cov(Y_{ik},Y_{i'k}|\mbG_{Sk}) = \sigma^2_u$. Then the ICC between $Y_{ik}$ and $Y_{i'k}$ conditional on $\mbG_{Sk}$ is 
\[
\rho_{Y} = \frac{\sigma^2_u}{\sigma^2_u+\sigma^2_e}.
\]
As a result, the variance of $\Y_{Sk}$ for $k$th network given $\mbG_{Sk}$ is $\Cov(\Y_{Sk}|\mbG_{Sk}) = \sigma^2_{Y} \cdot V_{Sk}$, where $V_{Sk} = (1-\rho_{Y})I_n+\rho_Y J_n$, $I_n$ is a $n \times n$ identity matrix, $J_n$ is a $n \times n$ matrix with all elements be 1.

Based on model \eqref{m2-delta}, the BLUE estimator of $\theta_{II}$ is given by generalized least squares
\[
\hat{\theta}_{II} = \left(\sum\limits^K_{k=1} \mbG'_{Sk}V^{-1}_{Sk} \mbG_{Sk}\right)^{-1}  \left(\sum\limits^K_{k=1} \mbG'_{Sk} \Y_{Sk}\right)
\]
with 
\[
\Var(\hat{\theta}_{II}) = \sigma^2_{Y} U^{-1}_{Sk}, 
\]
where $U_{Sk}=\sum\limits^K_{k=1} \mbG'_{Sk}V^{-1}_{Sk} \mbG_{Sk}$. When $K \to \infty$, $\sqrt{K}(\hat{\beta}-\beta)$ is asymptotic normal distributed as $N(0, \Sigma_S)$, where 
\[
\Sigma_S = \lim_{K \to \infty} \sigma^2_{Y} (U_{Sk}/K )^{-1}=
\sigma^2_{Y}\times \frac{1+(n-1)\rho_Y}{p(1-p)n} \times
\begin{bmatrix}
 p & -p \\
 -p & 1\\
\end{bmatrix}
\] 
and the derivation of $\Sigma_S$ is placed in Section \ref{Alter-var}.

\begin{figure}[!htb]
\centering

\minipage{0.8\textwidth}%
  \includegraphics[width=\linewidth]{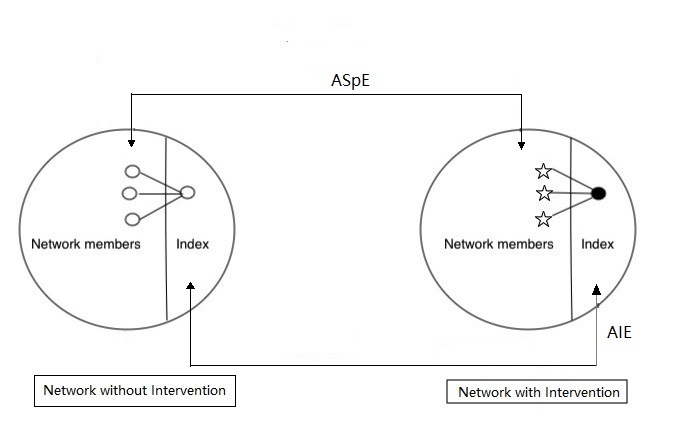}
  \caption{Egonetwork-based design and subsets of data used for the identification of each effect using models \eqref{m2-tau} and \eqref{m2-delta} in Section \ref{AlterModel-2} } \label{fig:design}
\endminipage
\end{figure}

\vspace{3mm}

\noindent \textbf{Remark}: Compare to the regression model proposed in main text (we call it single regression model here), the alternative regression model proposed here has more freedom to modelling the observed data since we can make different assumptions to $\epsilon_{1k}$ and $\epsilon_{ik}$. However, for estimating the AIE, the single regression model text exploits the entire individuals in networks without intervention. Thus, for estimating the AIE, the single regression model is more efficient than using \eqref{m2-tau} only.

\subsubsection{Sample Size and Power Calculation}

Based on models \eqref{m2-tau} and \eqref{m2-delta}, and the derived BLUE estimators, we report the hypothesis testing procedures for two hypotheses on the AIE and the ASpE. We then derive the required number of index participants to have adequate power to test the causal effects at a given significant level.

\textit{The AIE hypothesis test based on \eqref{m2-tau} ($HIE_2$)}. Here we want to test the hypothesis of no average individual effect, that is, $H_0: \tau=0.$ To test this hypothesis, we rely on the use of the two sided $Z-$test statistc: $Z_{\tau} = \sqrt{K}(\hat{\tau}/\sigma_{\tau})$, where $\sigma^2_{\tau} = \sigma^2_{Y_1}/\{p(1-p)\}$. Using the BLUE estimator of $\theta_I$, $Z_{\tau}$ follows a standard normal distribution given the null hypothesis. To against the null hypothesis, the alternative hypothesis, $H_1: \tau \neq 0$, represents that $Z_{\tau}$ dose not follow $N(0,1)$ under the alternative hypothesis. Assume that the effect size of $\tau$ is $\Delta_{\tau}$. Given a significant level of the test, $\alpha$, the probability to reject $H_0$ when it is true is $P(|Z_{\tau}|>z_{1-\alpha/2} \vert \tau = 0)$ with critical value $z_{1-\alpha/2}$. Then the power of the test is $\text{power}(\Delta_{\tau}) = P\{|\sqrt{K}(\hat{\tau}-\Delta_{\tau})/\hat{\sigma}_{\tau})|\geq  z_{1-\alpha/2} -\sqrt{K} \Delta_{\tau}/\sigma_{\tau}\vert Z_{\tau} \sim N(\Delta_{\tau},1)\} =1- \Phi(z_{1-\alpha/2}-\sqrt{K}\Delta_{\tau}/\sigma_{\tau})$, where $\Phi(\cdot)$ is the cumulative distribution function of the standard normal distribution and $z_{\lambda} = \Phi^{-1}(\lambda)$ for any $\lambda \in [0,1]$. Given the required power $1-\lambda$, at significant level $\alpha$, we solve $\text{power}(\Delta_{\tau})$ for $K$ to obtain the required number of index participant of $HIE_2$, $K_{\tau}$:
\begin{equation}
K_{\tau}=\frac{\sigma^2_{\tau}(z_{1-\alpha/2}+z_{1-\lambda})^2}{\Delta_{\tau}^2}  =  \frac{\sigma^2_{Y_1} (z_{1-\alpha/2}+z_{1-\lambda})^2}{p(1-p) \Delta_{\tau}^2} \label{m2-K-tau} % \tag{S.18}
\end{equation}

\textit{The ASpE hypothesis test based on \eqref{m2-delta} ($HSpE_2$)}. Here we focus on the hypothesis of no ASpE, $H_0: \delta=0$. To test this hypothesis, we use the two-sized Z-test statistic: $Z_{\delta} = \sqrt{K}(\hat{\delta}/\hat{\sigma}_{\delta})$, where
\[
\sigma^2_{\delta} = \frac{\sigma^2_{Y}\{1+(n-1)\rho_Y\}}{n p(1-p)}.
\]
Similar to $HIE_2$, $Z_{\delta}$ follows a standard normal distribution given $H_0$. Assuming the effect size of $\delta$ is $\Delta$, then given a significant level $\alpha$, the power of the test is $\text{power}(\Delta_{\delta}) =1- \Phi(z_{1-\alpha/2}-\sqrt{K}\Delta_{\delta}/\sigma_{\delta})$. We solve $\text{power}(\Delta_{\delta})$ for $K$ to obtain the required number of index participants to achieve a adequate power of the test, $1-\lambda$, given significance level, $\alpha$. To be specific,
\begin{equation}
K_{\delta}=\frac{\sigma^2_{\delta}(z_{1-\alpha/2}+z_{1-\lambda})^2}{\Delta_{\delta}^2} =  \frac{\sigma^2_{Y}\{1+(n-1)\rho_Y\} (z_{1-\alpha/2}+z_{1-\lambda})^2}{n p(1-p) \Delta_{\delta}^2}.  \label{m2-K-delta} %\tag{S.19}
\end{equation}

% When people interested in testing HDSE, the required number of networks are $\max(K_{\tau},K_{\delta})$.
The required number of index participants for testing the AIE and the ASpE simultaneously with power $1-\lambda$ is defined to be $\max(K_{\tau},K_{\delta})$ at significant level $\alpha$, which is the maximum value of $K_{\tau}$ and $K_{\delta}$. 

\subsubsection{Numerical Simulations}

From \eqref{m2-K-tau} and \eqref{m2-K-delta}, we observe that $K_{{\tau}}$ is related to the probability of index treatment $p$ and the effect size $\Delta_{\tau}$, while $K_{{\delta}}$ is related to $p$, the network size $n$, the intra-class correlation $\rho_Y$, and the effect size $\Delta_{\delta}$. In particular, we can see that smaller effect size $\Delta_{\tau}$ inflates the required number of index participants,
and $K_{\delta}$ increase with $\rho_{Y}$ and decrease with $n$.
smaller effect size $\Delta_{\delta}$ also inflates the required samples size of index participants for testing the ASpE.

\begin{figure}[!htb]
\centering

\minipage{0.7\textwidth}%
  \includegraphics[width=\linewidth]{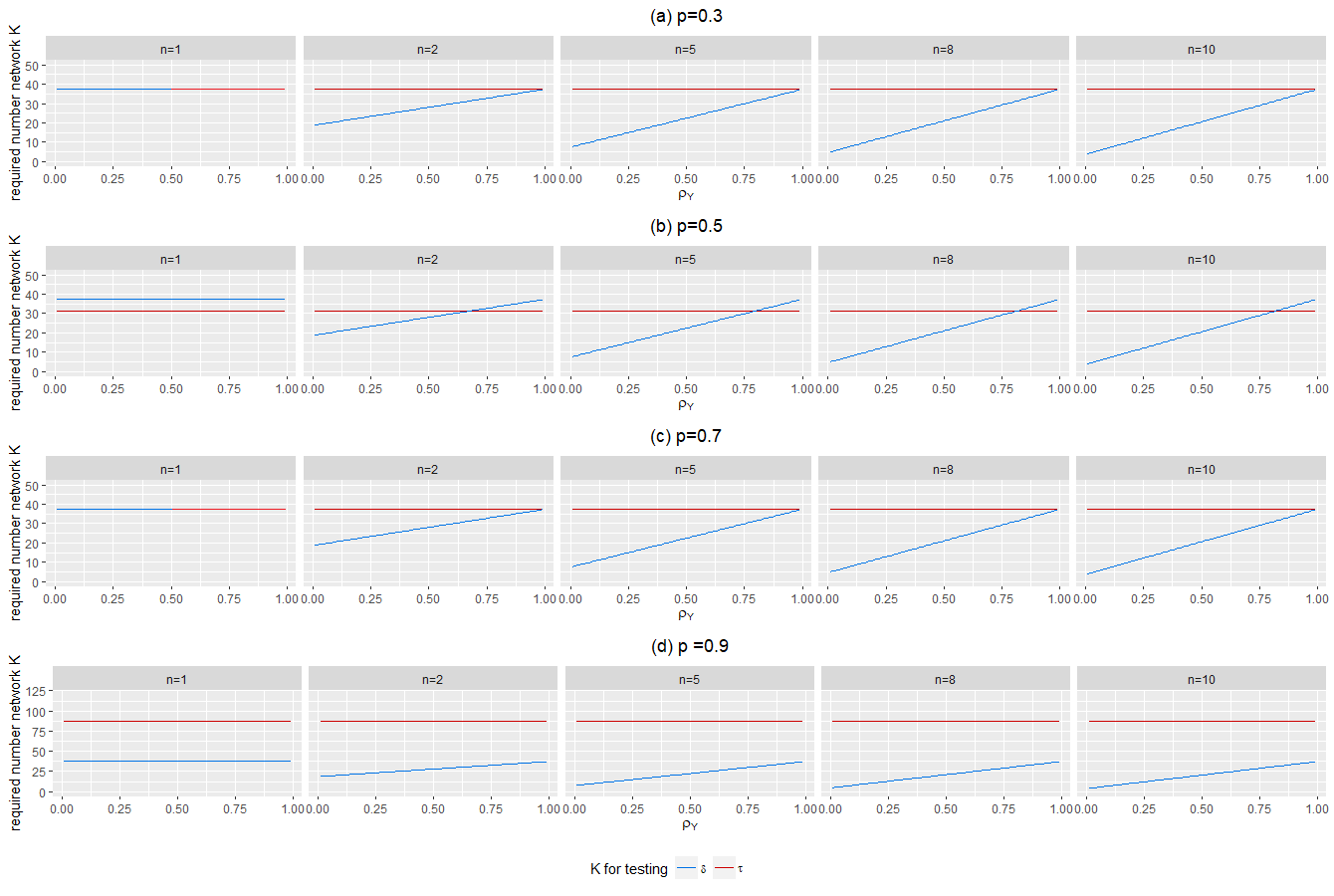}
  \caption{Required number of index participants for testing the AIE and ASpE given different value of $p$: $\Delta_{\tau}=\Delta_{\delta}=1$, $\sigma^2_{Y_1} = \sigma^2_Y=1$. The line with two colors represents that the two lines of $K_{\tau}$ and $K_{\delta}$ overlap with each other.} \label{fig:alt-design} 
\endminipage
\end{figure}

We fix $\sigma^2_{Y_1}$, $\sigma^2_Y$, $\Delta_{\tau}$ and $\Delta_{\delta}$ at 1, while we let $\rho_{Y}$ vary from 0 to 1, $n \in \{1,2,5,8,10\}$, and $p = c(0.3,0.5,0.7,0.9)$. Figure \ref{fig:alt-design} shows the required number of networks for $HIE_2$ and $HSpE_2$ with $\alpha = 5\%$ and $1-\lambda=80\%$.
As we can see, for each $p$, the patterns of $K_{\tau}$ and $K_{\delta}$ with $\rho_Y$ and $n$ are displayed as we expected. $K_{\tau}$ only changes with $p$, and the optimal value of $p$ is 0.5. Given $p$ and $n$, $K_{\delta}$ increases with $\rho_Y$. Smaller $n$ inflates $K_{\delta}$.

%Figure 3 highlights how the required number of index participants of different tests varies with $p$ given $n=5$. For any $\rho_{Y}$ and $n$, we notice that $p=0.5$ is the optimal value for testing the ASpE, the overall effect, and testing the AIE and the ASpE effect simultaneously, but we see different patterns for testing the AIE only, \textcolor{blue}{where $K_{\tau}$ has a larger increasing rate with one unit increases in $\rho_Y$,} and $p=0.7$ performs better than $p=0.5$ for smaller $\rho_Y$.

%As we can see, for each $p$, the patterns of $K_{\tau}$, $K_{\delta}$, $K_{\tau\delta}$, and $K_{o}$ with $\rho_Y$ and $n$ are displayed as we expected. Figure 3 highlights how the required number of index participants of different tests varies with $p$ given $n=5$. For any $\rho_{Y}$ and $n$, we notice that $p=0.5$ is the optimal value for testing the ASpE, the overall effect, and testing the AIE and the ASpE effect simultaneously, but we see different patterns for testing the AIE only, \textcolor{blue}{where $K_{\tau}$ has a larger increasing rate with one unit increases in $\rho_Y$,} and $p=0.7$ performs better than $p=0.5$ for smaller $\rho_Y$.

\subsubsection{Derivations of Variance for Alternative Regression-based Estimators \label{Alter-var}}

%\subsection{Derivations of $\Sigma_D$ and $\Sigma_S$}
In Section \ref{AlterModel-2}, we know that $\Var(\hat{\beta}_I) = \sigma^2_{Y_1}U^{-1}_{Dk}$, where
\[
U_{Dk} = \Z'_1 \Z_1 = \begin{bmatrix}
 K & \sum\limits^{K}_{k=1} Z_{1k} \\
 \sum\limits^{K}_{k=1} Z_{1k} & \sum\limits^{K}_{k=1} Z^2_{1k}
\end{bmatrix}.
\]
We define $U_D = \lim_{K \to \infty} U_{Dk}/K$, then
\[
U_D =  \lim_{K \to \infty} \frac{1}{K} \begin{bmatrix}
 K & \sum\limits^{K}_{k=1} Z_{1k} \\
 \sum\limits^{K}_{k=1} Z_{1k} & \sum\limits^{K}_{k=1} Z^2_{1k}
\end{bmatrix} = \begin{bmatrix}
 1 & p \\
p & p
\end{bmatrix},
\]
where we applied the properties: $\lim_{K \to \infty}  \sum\limits^{K}_{k=1} Z_{1k}/K =p $ and  $\lim_{K \to \infty} \sum\limits^{K}_{k=1} Z^2_{1k}/K =p $.
As a result, 
\[
\Sigma_D = \lim_{K \to \infty} \sigma^2_{Y_1} (U_{Dk}/K)^{-1} = \sigma^2_{Y_1} U^{-1}_D = \frac{\sigma^2_{Y_1}}{p(1-p)} \begin{bmatrix}
 1 & p \\
p & p
\end{bmatrix}.
\]

Similarly, we have  $\Var(\hat{\beta}_{II}) = \sigma^2_{Y}U^{-1}_{Sk}$ with
\[
\begin{aligned}
U_{Sk} &= \sum\limits^{K}_{k=1} \mbG'_{Sk}V^{-1}_{Sk} \mbG_{Sk} = cS_k +dT_k,
\end{aligned}
\]
where $S_k = \sum\limits^{K}_{k=1}\mbG'_{Sk} \mbG_{Sk}$ and $T_k =  \sum\limits^{K}_{k=1}\mbG'_{Sk}J_n \mbG_{Sk}  $.
Then with the properties: $\lim_{K \to \infty}  \sum\limits^{K}_{k=1} G_{ik}/K =p$ and  $\lim_{K \to \infty}  \sum\limits^{K}_{k=1} G^2_{ik}/K =p$ 
\[
\begin{aligned}
S &= \lim_{k \to \infty} \frac{1}{K} S_k =  \lim_{k \to \infty} \frac{1}{K} \sum\limits^{K}_{k=1}\mbG'_{Sk} \mbG_{Sk} \\
 & = \lim_{k \to \infty} \frac{1}{K} \sum\limits^{K}_{k=1} \begin{bmatrix}
  n & \sum\limits^{n+1}_{i=2} G_{ik} \\
 =    \sum\limits^{n+1}_{i=2} G_{ik} &  \sum\limits^{n+1}_{i=2} G^2_{ik}
 \end{bmatrix} \\
 &= \begin{bmatrix}
  n & np \\
  n p &  np
 \end{bmatrix} \\
\end{aligned}
\]
and
\[
\begin{aligned}
T &= \lim_{k \to \infty} \frac{1}{K} T_k =  \lim_{k \to \infty} \frac{1}{K} \sum\limits^{K}_{k=1}\mbG'_{Sk} J_n \mbG_{Sk} \\
 & = \lim_{k \to \infty} \frac{1}{K} \sum\limits^{K}_{k=1} \begin{bmatrix}
  n^2 & n \sum\limits^{n+1}_{i=2} G_{ik} \\
  n  \sum\limits^{n+1}_{i=2} G_{ik} &  (\sum\limits^{n+1}_{i=2} G_{ik})  (\sum\limits^{n+1}_{i=2} G_{ik})
 \end{bmatrix} \\
 &  = \begin{bmatrix}
  n^2 & n^2p \\
  n^2 p &  n^2 p
 \end{bmatrix}.
\end{aligned}
\]
We define $U_S =  \lim_{K \to \infty}U_{Sk}/K$
and recall that 
$c = 1/(1-\rho_Y)$ and
$d=-\rho_Y/[1-\rho_Y\{1+(n-1)\rho_Y\}]$.
Then 
\[\begin{aligned}
U_S &=  \lim_{K \to \infty} cS_k +dT_k = cS+dT \\
&  = \left( \frac{n}{1-\rho_Y} -\frac{\rho_Yn^2}{(1-\rho_Y)\{1+(n-1)\rho_Y\}}\right) \begin{bmatrix}
  1 & p \\
   p &  p
 \end{bmatrix} 
\end{aligned}
\]
and
\[
\begin{aligned}
U^{-1}_S &= \left( \frac{n}{1-\rho_Y} -\frac{\rho_Yn^2}{(1-\rho_Y)\{1+(n-1)\rho_Y\}}\right)^{-1} \frac{1}{ p(1- p)}\begin{bmatrix}
p & -p \\
  - p &  1
 \end{bmatrix} \\
 & = \frac{1+(n-1)\rho_Y}{n} \frac{1}{ p(1- p)}\begin{bmatrix}
p & -p \\
  - p &  1
 \end{bmatrix}.
\end{aligned}
\]
As a result, we have 
\[
\begin{aligned}
\Sigma_S &= \lim_{K \to \infty}\sigma^2_Y(U_{Sk}/K)^{-1} \\
&  = \sigma^2_Y U^{-1}_S \\
& = \sigma^2_Y  \frac{1+(n-1)\rho_Y}{np(1- p)}\begin{bmatrix}
p & -p \\
  - p &  1
 \end{bmatrix}.
\end{aligned}
\]

%\iffalse  %\iftrue

%\bibliography{ref}

%Put your short appendix here.  Remember, longer appendices are
%possible when presented as Supplementary Web Material.  Please 
%review and follow the journal policy for this material, available
%under Instructions for Authors at \texttt{http://www.biometrics.tibs.org}.

%\label{lastpage}

\end{document}